\documentclass[12pt]{article}
\usepackage{a4wide,epsfig,amsmath,amssymb,cite,scalefnt,graphicx}
\usepackage{slashed}
\usepackage{axodraw4j}
\usepackage{pstricks}
\usepackage{color}
\usepackage{changepage}
\usepackage{tabularx}
\usepackage{graphics}
\usepackage{array}

\begin{document}

\def \inparg{\leftskip = 40pt\rightskip = 40pt} \def \outparg{\leftskip = 0 pt\rightskip = 0pt} \def\bolone{{\bf 1}} \def\Ocal{{\cal O}} \def\NSVZ{{\rm NSVZ}} \def\DR{{\rm DRED}} \def\DG{{\rm DREG}} \def\DREDp{{\rm DRED}'} \def\npb{{\it{Nucl.\ Phys.}\ }{\bf B}} 

\def\prl{Phys.\ Rev.\ Lett.\ }
\def\zpc{Z.\ Phys.\ {\bf C}}
\def\msbar{{\overline{\rm MS}}}
\def\drbar{{\overline{\rm DR}}}

\newcommand{\third}{\mbox{\small{$\tfrac{1}{3}$}}}
\newcommand{\pitwo}{\mbox{\small{$\tfrac{\pi}{2}$}}}
\newcommand{\pisix}{\mbox{\small{$\tfrac{\pi}{6}$}}}
\newcommand{\MSbar}{\overline{\mbox{MS}}}
\newcommand{\MOMs}{\mbox{\footnotesize{MOM}}}

\def\tlambda{\tilde \lambda}
\def\gbar{\overline{g}}
\def\ttheta{\tilde \theta}
\def\tTheta{\tilde \Theta}
\def\tLambda{\tilde \Lambda}
\def\yhat{\hat y}
\def\that{\hat t}
\def\Gtil{\hat G}
\def\tZ{\tilde Z}
\def\tz{\tilde z}
\def\ov #1{\overline{#1}}
\def\thetabar{{\overline\theta}}
\def\alphadot{\dot\alpha}
\def\betadot{\dot\beta}
\def\Qbar{\overline Q}
\def\cbar{\overline c}
\def\fbar{\overline f}
\def\Fbar{\overline F}
\def\Lambdabar{\overline \Lambda}
\def\psibar{\overline{\psi}}
\def\betabar{\overline{\beta}}
\def\wt #1{\widetilde{#1}}
\def\Tr{\mathop{\rm Tr}}
\def\det{\mathop{\rm det}}
\def\wtl{\widetilde{\lambda}}
\def\wh{\widehat}
\def\th{{\theta}}
\def\bth{{\overline{\theta}}}
\def\eps{\epsilon}
\def\atil{\tilde\alpha}
\def\frak#1#2{{\textstyle{{#1}\over{#2}}}}
\def\nhalf{${\cal N} = \frak{1}{2}$}
\def\none{${\cal N} = 1$}
\def\ntwo{${\cal N} = 2$}

\def\ybar{\overline y}
\def\mbar{\overline m}
\def\tautil{\tilde\tau}
\def\chibar{\overline\chi}
\def\nutil{\tilde\nu}
\def\mutil{\tilde\mu}
\def\rhotil{\tilde\rho}
\def\sigtil{\tilde\sigma}
\def\gatil{\tilde\ga}
\def\Btil{\tilde B}
\def\Bbar{\overline B}
\def\Ttil{\tilde T}
\def\fhat{\hat f}
\def\Ahat{\hat A}
\def\Chat{\hat C}
\def\mbar{\overline{m}}
\def\etabar{\overline\eta}
\def\mubar{\overline{\mu}}
\def\deltabar{\overline\delta}
\def\alphabar{\overline\alpha}
\def\half{{\textstyle{{1}\over{2}}}}
\def\frak#1#2{{\textstyle{{#1}\over{#2}}}}
\def\frakk[#1#2{{{#1}\over{#2}}}
\def\go{\rightarrow}
\def\lambdabar{\overline\lambda}
\def\lambdahatbar{\overline{\hat\lambda}}
\def\lambdahat{\hat\lambda}
\def\Dtil{\tilde D}
\def\Dhat{\hat D}
\def\Dbar{\overline D}
\def\Ftil{\tilde F}
\def\Fhat{\hat F}
\def\Atil{\tilde A}
\def\sigmabar{\overline\sigma}
\def\phibar{\bar\phi}
\def\phitilbar{\overline{\tilde{\phi}}}
\def\psitilbar{\overline{\tilde{\psi}}}
\def\Phibar{\overline\Phi}
\def\psibar{\overline\psi}
\def\Fbar{\overline F}
\def\TeV{{\rm TeV}}
\def\GeV{{\rm GeV}}
\def\Ghat{\hat\Gamma}
\def\Btil{\tilde B}
\def\Dtil{\tilde D}
\def\Etil{\tilde E}
\def\Ttil{\tilde T}
\def\Ytil{\tilde Y}
\def\Qtil{\tilde Q}
\def\Ltil{\tilde L}
\def\Atil{\tilde \alpha}
\def\ctil{\tilde c}
\def\dtil{\tilde \delta}
\def\etil{\tilde \epsilon}
\def\gtil{\tilde \gamma}
\def\mtil{\tilde \mu}
\def\ntil{\tilde \nu}
\def\rtil{\tilde \rho}
\def\stil{\tilde \sigma}
\def\xtil{\tilde \xi}
\def\ztil{\tilde \zeta}
\def\ttil{\tilde t}
\def\util{\tilde u}
\def\phitil{\tilde\phi}
\def\psitil{\tilde\psi}
\def\Ncal{{\cal N}}
\def\Ftil{\tilde F}
\def\Ytil{\tilde Y}
\def\alphadot{\dot\alpha}
\def\betadot{\dot\beta}
\def\deltadot{\dot\delta}
\def\Vhat{\hat V}
\def\Rhat{\hat R}
\def\Abar{\overline A}
\def\Bbar{\overline B}
\def\Mbar{\overline M}
\def\Nbar{\overline N}
\def\Hbar{\overline H}
\def\ahat{\hat a}
\def\bhat{\hat b}
\def\sy{supersymmetry}
\def\sic{supersymmetric}
\def\pa{\partial}
\def\pabar{\overline\partial}
\def\smgroup{$SU_3\otimes\ SU_2\otimes\ U_1$} \def\stw{\sin^2\th_W}

\input amssym.def
\input amssym
\baselineskip 14pt
\parskip 6pt

\def\npb{Nucl. Phys. B}
\def\plb{Phys. Lett. B}
\def\pa{\partial}
\def\be{\begin{equation}}
\def\ee{\end{equation}}
\def\bea{\begin{align}}
\def\eea{\end{align}}
\def\nn{\nonumber\\}
\def\n{\nonumber}
\def\tr{\hbox{tr}}
\def\trhat{\hat\tr}
\def\barh{\overline h}
\def \de{\delta}
\def \De{\Delta}
\def \si{\sigma}
\def \Ga{\Gamma}
\def \ga{\gamma}
\def \nab{\nabla}
\def \pr{\partial}
\def \d{{\rm d}}
\def \tr{{\rm tr }}
\def \ta{{\tilde a}}
\def \hs{\hat s}
\def \hr{\hat r}
\def \br{\bar r}
\def \hG{{\hat G}}
\def \bI{{\bar I}}
\def \bL{{\bar L}}
\def \bT{{\bar T}}
\def \bO{{\bar O}}
\def \by{{\bar y}}
\def\bx{{\bar x}}
\def\ba{{\bar a}}
\def\bb{{\bar b}}
\def\bh{{\bar h}}
\def\bm{{\bar m}}
\def \bY{{\bar Y}}
\def \bW{{\bar W}}
\def \bQ{{\bar Q}}
\def \bl{{\lambda}}
\def \bpsi{{\bar \psi}}
\def \bsi{{\bar \sigma}}
\def \bchi{{\bar \chi}}
\def \bphi{{\bar \phi}}
\def \bta{{\bar \eta}}
\def\ba{{\bar a}}
\def\bb{{\bar b}}
\def \rO{{\rm O}}
\def \l{\big \langle}
\def \r{\big \rangle}
\def \ep{\epsilon}
\def \vep{\varepsilon}
\def \half{{\textstyle {1 \over 2}}}
\def \thir{{\textstyle {1 \over 3}}}
\def \quar{{\textstyle {1 \over 4}}}
\def \ts{\textstyle}
\def\del{{\rm d}}
\def \A{{\cal A}}
\def \B{{\cal B}}
\def \C{{\cal C}}
\def \D{{\cal D}}
\def \E{{\cal E}}
\def \F{{\cal F}}
\def \G{{\cal G}}
\def \H{{\cal H}}
\def \I{{\cal I}}
\def \J{{\cal J}}
\def \K{{\cal K}}
\def \L{{\cal L}}
\def \M{{\cal M}}
\def \N{{\cal N}}
\def \O{{\cal O}}
\def \P{{\cal P}}
\def \Q{{\cal Q}}
\def \R{{\cal R}}
\def \S{{\cal S}}
\def \T{{\cal T}}
\def \V{{\cal V}}
\def \X{{\cal X}}
\def \Y{{\cal Y}}
\def \Z{{\cal Z}}
\def\fd{{\frak d}}
\def\fe{{\frak e}}
\def\fN{{\frak r}}
\def \tb{{\tilde {\rm b}}}
\def \tx{{\tilde {\rm x}}}
\def \ty{{\tilde {\rm y}}}
\def \tK{{\tilde K}}
\def \tsi{{\tilde \sigma}}
\def \h{{\rm h}}
\def \a{{\rm a}}
\def \b{{\rm b}}
\def \d{{\rm d}}
\def \e{{\rm e}}
\def \x{{\rm x}}
\def \y{{\rm y}}
\def\uF{\bar{F}}
\def\uG{\bar{G}}
\def\uA{\bar{A}}
\def\uB{\bar{B}}
\def\uC{\bar{C}}
\def\uD{\bar{D}}
\def\uE{\bar{E}}
\def\uH{\bar{H}}
\def\ual{\bar{\alpha}}
\def\ube{\bar{c}}
\def\uga{\bar{\gamma}}
\def\ude{\bar{\delta}}
\def\uet{\bar{\eta}}
\def\uep{\bar{\epsilon}}
\def\hDel{\hat \Delta}
\def\wL{{\widetilde \L}}
\def\hrho{{\tilde \rho}}
\def\bxi{{\bar \xi}}
\def\blam{{\bar \lambda}}
\def \vphi{{\varphi}}
\def \tD{{\tilde D}}
\def\hbet{c}
\def\hhbet{{\hat c}}
\def\hB{{\hat B}}
\def\btil{\tilde c}
\def\cirk{\,{\raise1pt \hbox{${\scriptscriptstyle \circ}$}}\,} \def\limu#1{\mathrel{\mathop{\sim}\limits_{\scriptscriptstyle{#1}}}}
\def\toinf#1{\mathrel{\mathop{\longrightarrow}\limits_{\scriptstyle{#1}}}}
\def \olr{{\raise6.5pt\hbox{$\leftrightarrow  \! \! \! \! \!$}}}

\font\ninerm=cmr9 \font\ninesy=cmsy9
\font\eightrm=cmr8 \font\sixrm=cmr6
\font\eighti=cmmi8 \font\sixi=cmmi6
\font\eightsy=cmsy8 \font\sixsy=cmsy6
\font\eightbf=cmbx8 \font\sixbf=cmbx6
\font\eightit=cmti8
\def\eightpoint{\def\rm{\fam0\eightrm}
  \textfont0=\eightrm \scriptfont0=\sixrm \scriptscriptfont0=\fiverm
  \textfont1=\eighti  \scriptfont1=\sixi  \scriptscriptfont1=\fivei
  \textfont2=\eightsy \scriptfont2=\sixsy \scriptscriptfont2=\fivesy
  \textfont3=\tenex   \scriptfont3=\tenex \scriptscriptfont3=\tenex
  \textfont\itfam=\eightit  \def\it{\fam\itfam\eightit}%
  \textfont\bffam=\eightbf  \scriptfont\bffam=\sixbf
  \scriptscriptfont\bffam=\fivebf  \def\bf{\fam\bffam\eightbf}%
  \normalbaselineskip=9pt
  \setbox\strutbox=\hbox{\vrule height7pt depth2pt width0pt}%
  \let\big=\eightbig  \normalbaselines\rm}
\catcode`@=11 %
\def\eightbig#1{{\hbox{$\textfont0=\ninerm\textfont2=\ninesy
  \left#1\vbox to6.5pt{}\right.\n@@space$}}} \def\vfootnote#1{\insert\footins\bgroup\eightpoint
  \interlinepenalty=\interfootnotelinepenalty
  \splittopskip=\ht\strutbox %
  \splitmaxdepth=\dp\strutbox %
  \leftskip=0pt \rightskip=0pt \spaceskip=0pt \xspaceskip=0pt
  \textindent{#1}\footstrut\futurelet\next\fo@t}
\catcode`@=12 %
\def\today{\number\day\ \ifcase\month\or January\or February\or March\or April\or May\or June\or July\or August\or September\or October\or November\or December\fi, \number\year} \def\now{\number\time} \font \bigbf=cmbx10 scaled \magstep1

\input epsf

\numberwithin{equation}{section}

\begin{titlepage}
\begin{flushright}
LTH1045\\
\end{flushright}
\date{}
\vspace*{3mm}

\begin{center}
{\Huge The $a$-function in six dimensions}\\[12mm] {\bf J.A.~Gracey, I.~Jack\footnote{{\tt dij@liv.ac.uk}} and C.~Poole\footnote{{\tt c.poole@liv.ac.uk}}}\\ 

\vspace{5mm}
Dept. of Mathematical Sciences,
University of Liverpool, Liverpool L69 3BX, UK\\

\end{center}

\vspace{3mm}
\begin{abstract}
The $a$-function is a proposed quantity defined in even dimensions which has a monotonic behaviour along RG flows, related to the $\beta$-functions via a gradient flow equation.
We study the $a$-function for a general scalar theory in six dimensions, using the $\beta$-functions up to three-loop order for both the $\MSbar$ and MOM schemes (the latter presented here for the first time at three loops).
\end{abstract}

\vfill

\end{titlepage}


\section{Introduction}
There has been considerable recent interest over the possibility (first raised by Cardy\cite{Cardy}) of a four-dimensional generalisation of  Zamolodchikov's $c$-theorem\cite{Zam} in two dimensions; namely a function $a(g)$ of the couplings $g^I$  which has monotonic behaviour under renormalisation-group (RG) flow or which is defined at fixed points such that $a_{\rm UV}-a_{\rm IR}>0$. These two possibilities are referred to as the strong or weak $a$-theorem, respectively.
A proof of the weak $a$-theorem has been proposed by Komargodski and Schwimmer\cite{KS} with further analysis  in Ref.\cite{Luty}.

The strong $a$-theorem has been proved valid for small values of the couplings\cite{Analog,Analoga}, using Wess-Zumino consistency conditions for the response of the theory defined on curved spacetime, and with $x$-dependent couplings $g^I(x)$, to a Weyl rescaling of the metric\cite{Weyl}.
 A function $A$ is defined which satisfies the crucial equation
\be \pa_IA =T_{IJ}\beta^J\, ,
\label{grad}
\ee
 for a function $T_{IJ}$ which is defined in terms of RG quantities and may  in principle be computed perturbatively within the theory extended to curved spacetime and $x$-dependent $g^I$.
Eq.~(\ref{grad}) implies
\be
\mu \frac{d}{d\mu} A=\beta^I\frac{\pa}{\pa g^I} A=G_{IJ} \beta^I\beta^J
\ee
where $G_{IJ}=T_{(IJ)}$; thus verifying the strong $a$-theorem so long as $G_{IJ}$ is positive-definite, a property which holds at least  for weak couplings in four dimensions.
It is clear that if
$A$ satisfies an equation of the form Eq.~\eqref{grad} then so does
\be
A'=A+g_{IJ}\beta^I\beta^J
\label{free}
\ee
for any $g_{IJ}$ (for a different $T_{IJ}$, of course).

Further extensions of this general framework have been explored in Refs.~\cite{Other, KZ, KZa}. We should
mention explicitly here that for theories with a global symmetry, $\beta^I$ in these equations should be replaced by a $B^I$ which is defined, for instance, in Ref.\cite{Analoga}. However, it was shown in Refs.\cite{Fortin, FortinC} that the two quantities only begin to differ at three loops for theories in four dimensions.

It was shown in Ref.~\cite{GrinsteinCKA} that equations of a similar form to the above  may be derived (in a similar manner) for a renormalisable theory in six dimensions\footnote{ See Ref.~\cite{Nakthree} for an analogous extension to three dimensions; and Ref.~\cite{jjp} for a recent explicit construction of an $a$-function in three dimensions. See also Ref.~\cite{elvang} for attempts to derive a weak $a$-theorem in six and general $d$ dimensions using the methods of Ref.~\cite{KS}. }; though the definition of $A$ and $T_{IJ}$ as renormalisation-group quantities is of course different. For instance,
\be
A=6a+ b_1-\tfrac{1}{15} b_3+W_I\beta^I
\ee
where $W_I$, like $T_{IJ}$, has a definition  in terms of RG quantities, $a$ is the $\beta$-function corresponding to the six-dimensional Euler density
\be E_6=\epsilon_{\mu_1\mu_2\mu_3\mu_4\mu_5\mu_6}
\epsilon_{\nu_1\nu_2\nu_3\nu_4\nu_5\nu_6}R^{\mu_1\mu_2\nu_1\nu_2}
R^{\mu_3\mu_4\nu_3\nu_4}R^{\mu_5\mu_6\nu_5\nu_6}
\ee
and $b_1$ and $b_3$ are the $\beta$-functions corresponding respectively to
\be
 L_1=-\tfrac{1}{30}K_1+\tfrac14K_2-K_6, \quad
L_3=-\tfrac{37}{6000}K_1+\tfrac{7}{150}K_2-\tfrac{1}{75}K_3+\tfrac{1}{10}K_5
+\tfrac{1}{15}K_6,
\ee
where (following the notation of Ref.~\cite{GrinsteinCKA})
\begin{align}
 K_1=R^3,\quad K_2=RR^{\kappa\lambda}R_{\kappa\lambda},&\quad
K_3=RR^{\kappa\lambda\mu\nu}R_{\kappa\lambda\mu\nu},\quad
K_4=R^{\kappa\lambda}R_{\lambda\mu}R^{\mu}{}_{\kappa},\nn
K_5=R^{\kappa\lambda}R_{\kappa\mu\nu\lambda}R^{\mu\nu},&\quad
K_6=R^{\kappa\lambda}R_{\kappa\mu\nu\rho}R_{\lambda}{}^{\mu\nu\rho}.
\end{align}
We use the notation $A$ to avoid confusion with the $\tilde a$ of Ref.~\cite{GrinsteinCKA} which differs by a factor of 6.
In six dimensions, $G_{IJ}$ has recently been computed to be negative definite at leading order for a multiflavor $\phi^3$ theory\cite{Sternew,Stertwo}. The six-dimensional case has also been considered in more general terms in Ref.~\cite{OsbSter}.
Our purpose here is to extend the results of Ref.~\cite{Sternew,Stertwo} to higher orders, again for the multiflavor $\phi^3$ theory. However, we shall do this by using the $\beta$-functions together with Eq.~\eqref{grad} to construct the quantities $A$ and $T_{IJ}$ order by order (rather than by a direct perturbative computation).  We shall compute the function $A$ up to 5 loop order in the standard $\MSbar$ renormalisation scheme, requiring a knowledge of the three-loop 
$\MSbar$ $\beta$-function. We shall find that a solution for $A$ and $T_{IJ}$ is only possible if the $\beta$-function coefficients satisfy a set of consistency conditions, and we shall be able to show that these conditions are invariant under the coupling redefinitions which must relate any pair of renormalisation schemes. We illustrate the redefinition process using the example of the MOM (momentum subtraction) scheme. We accordingly present for the first time the three-loop $\beta$-functions for MOM (the three-loop $\MSbar$
$\beta$-functions may be read off from the results presented in Ref.~\cite{kleb}, although they were not written down explicitly there\footnote{Earlier three-loop results were presented in Ref.~\cite{jag1,jag2}, but these do not determine the results for the general theory unambiguously, as we shall explain later.}). We shall also provide full details of the three-loop calculation for the $\MSbar$ $\beta$-functions, and then give a precise definition of the MOM scheme, explaining how the calculation may be adapted for this case. In the general case we shall be considering, the theory has a global symmetry; and just as in four dimensions, we shall find that at three loops the consistency conditions can only be satisfied if we replace $\beta^I$ by the quantity $B^I$ defined in Ref.~\cite{Analoga}.

The layout of the paper is as follows: in Section 2 we present the one and two loop results for the $\beta$-functions and also the lowest-order results for $A$ and $T_{IJ}$. In Section 3 we give an explanation of our computational methods and how they may be applied to the computation of the $\beta$-functions in both the $\MSbar$ and MOM schemes, and then go on to list the results for the three-loop $\MSbar$ $\beta$-function. In Section 4 we present our results for $A^{(5)}$ together with consistency conditions on the $\beta$-function coefficients, which must be satisfied in any scheme in order for Eq.~\eqref{grad} to hold.
In Section 5 we discuss the implementation of renormalisation scheme changes in general terms and then go on to focus on the case of the MOM scheme. We present our concluding remarks in Section 6, and finally the explicit three-loop MOM $\beta$-function together with some calculational details are given in the Appendices.

\section{One and two-loop results}
We consider the theory
\be
L=\tfrac12\pa_{\mu}\phi^i\pa^{\mu}\phi^i+\tfrac{1}{3!}g^{ijk}\phi^i\phi^j\phi^k,
\label{lag}
\ee
involving a multiplet of fields $\phi^i$ coupled via the tensor $g^{ijk}$.
The one and two loop $\beta$-functions are given by
\begin{align}
\beta^{(1)}&=~-g_{(1a)}+\tfrac{1}{12}g_{(1A)},\nn
\beta^{(2)}&=~c_{(2B)}g_{(2B)}+c_{(2C)}g_{(2C)}+c_{(2b)}g_{(2b)}+c_{(2c)}g_{(2c)}+c_{(2d)}g_{(2d)},
\label{betatwo}
\end{align}
where the tensor structures are defined by
\begin{align}
g_{(1a)}^{ijk}=~g^{ilm}g^{jmn}g^{knl},\quad g_{(2b)}^{ijk}&=g^{jpq}g^{kpr}g_{(1a)}^{iqr},\nn
g_{(2c)}^{ijk}=g^{ipr}g^{jpq}g^{qs}_{1A}g^{ksr},\quad
g_{(2d)}^{ijk}&=g^{imn}g^{jpq}g^{krs}g^{nqs}g^{mpr},
\label{tensa}
\end{align}
with also
\be
g_{(1A)}^{ij}=g^{ikl}g^{jkl}, \quad g_{(2B)}^{ij}=g^{ipq}g_{(1a)}^{jpq},\quad 
g_{(2C)}^{ij}=g^{imn}g^{jmq}g_{(1A)}^{nq}.
\label{tensc}
\ee
We also define 3-index quantities corresponding to the 2-index quantities of Eq.~\eqref{tensc} by
\be
g_{(1A)}^{ijk}=g_{(1A)}^{il}g^{ljk},\quad \hbox{etc.}
\label{anomdim}
\ee
We have therefore given the same label to both a two-index and a three-index tensor, but we hope that it will always be clear from the context which is meant. For structures which are not three-fold symmetric, we list one symmetrisation but it is to be understood that it is accompanied by its two symmetrised partners. We shall also, wherever possible, suppress indices as we have done in Eq.~\eqref{betatwo}. We display the tensor structures appearing in Eq.~\eqref{betatwo} in Table~\ref{fig1} (in which the index $i$ is always at the right, except for $g_{(2d)}$, which is completely symmetric in $i$, $j$ and $k$).

\begin{table}[h]
	\setlength{\extrarowheight}{0cm}
	\setlength{\tabcolsep}{24pt}
	\hspace*{-4cm}
	\centering
	\resizebox{8cm}{!}{
		\begin{tabular*}{20cm}{ccccc}
			\begin{picture}(103,111) (299,-247)
			\SetWidth{1.0}
			\SetColor{Black}
			\Arc(336,-193)(35.777,153,513)
			\Line(371,-193)(401,-193)
			\Line(316,-163)(300,-137)
			\Line(316,-223)(303,-246)
			\end{picture}
			&
			\resizebox{6cm}{!}{\begin{picture}(104,60) (298,-285)
				\SetWidth{0.7}
				\SetColor{Black}
				\Line(379,-252)(401,-252)
				\Line(315,-252)(299,-226)
				\Line(315,-253)(302,-276)
				\Arc(359,-252)(20,233,593)
				\Line(316,-252)(338,-252)
				\end{picture}
			}
			&
			&
			&
			\\
			{\Huge $g_{(1a)}$}
			&
			{\Huge $g_{(1A)}$}
			&
			&
			&
			\\
			&
			&
			&
			&
			\\
			\resizebox{6cm}{!}{\begin{picture}(104,60) (298,-285)
				\SetWidth{0.7}
				\SetColor{Black}
				\Line(379,-252)(401,-252)
				\Line(315,-252)(299,-226)
				\Line(315,-253)(302,-276)
				\Arc(359,-252)(20,233,593)
				\Line(316,-252)(338,-252)
				\Line(359,-232)(359,-271)
				\end{picture}}
			&
			\resizebox{6cm}{!}{\begin{picture}(104,60) (298,-285)
				\SetWidth{0.7}
				\SetColor{Black}
				\Line(379,-252)(401,-252)
				\Line(315,-252)(299,-226)
				\Line(315,-253)(302,-276)
				\Arc(359,-252)(20,233,593)
				\Line(316,-252)(338,-252)
				\Arc(359,-235.958)(13.042,-175.419,-4.581)
				\end{picture}}
			&
			\begin{picture}(103,111) (299,-247)
			\SetWidth{1.0}
			\SetColor{Black}
			\Arc(336,-193)(35.777,153,513)
			\Line(371,-193)(401,-193)
			\Line(316,-163)(300,-137)
			\Line(316,-223)(303,-246)
			\Line(336,-158)(336,-228)
			\end{picture}
			&
			\begin{picture}(103,111) (299,-247)
			\SetWidth{1.0}
			\SetColor{Black}
			\Arc(336,-193)(35.777,153,513)
			\Line(371,-193)(401,-193)
			\Line(316,-163)(300,-137)
			\Line(316,-223)(303,-246)
			\Arc[clock](306.976,-193)(20.024,92.794,-92.794)
			\end{picture}
			&
			\resizebox{3cm}{!}{\begin{picture}(66,82) (335,-223)
				\SetWidth{0.7}
				\SetColor{Black}
				\Line(368,-142)(368,-158)
				\Line(368,-158)(336,-222)
				\Line(368,-158)(400,-222)
				\Line(356,-182)(391,-203)
				\Line(346,-203)(364,-192)
				\Line(374,-186)(380,-182)
				\end{picture}}
			\\
			{\Huge $g_{(2B)}$}
			&
			{\Huge $g_{(2C)}$}
			&
			{\Huge $g_{(2b)}$}
			&
			{\Huge $g_{(2c)}$}
			&
			{\Huge $g_{(2d)}$}			
		\end{tabular*}
	}
	\caption{Tensor structures appearing in the one- and two-loop $\beta$-functions}
	\label{fig1}	
\end{table}

The coefficients in Eq.~\eqref{betatwo} are given in  $\MSbar$  by
\be
c_{(2B)}=\tfrac{1}{18},\quad
c_{(2C)}=-\tfrac{11}{432}, \quad c_{(2b)}=-\tfrac14,\quad c_{(2c)}=\tfrac{7}{72}, \quad c_{(2d)}=-\tfrac12.
\label{cmstwo}
\ee
Here and elsewhere we suppress a factor of $(64\pi^3)^{-1}$ for each loop order.

It was shown in Refs~\cite{Sternew,Stertwo} that Eq.~\eqref{grad} was valid at leading order
with
\begin{align} A^{(3)}=-\tfrac{1}{4}\lambda g^{ijk}(g_{(1a)}^{ijk}-\tfrac14g_{(1A)}^{ijk})&=-\tfrac{1}{4}\lambda(g^{klm}g^{knp}g^{lpq}g^{mqn}
-\tfrac14g_{(1A)}^{mn}g_{(1A)}^{mn}),\nn
T^{(2)}_{IJ}&=G^{(2)}_{IJ}=\lambda\delta_{IJ},
\label{Athree}
\end{align}
and
\be
\lambda=-\tfrac{1}{3240}.
\label{Athreea}
\ee
As mentioned earlier, our definition of $A$ differs from that of $\tilde a$  in Ref.~\cite{GrinsteinCKA} by a factor of $6$, introduced for convenience. As the notation implies, $G^{(2)}_{IJ}$ would require a two-loop perturbative calculation using the methods of Ref.~\cite{Analog,Analoga}, which was performed explicitly in Refs.~\cite{Sternew,Stertwo}; $A^{(3)}$ would correspondingly require a three-loop calculation, but its value was inferred by imposing Eq.~\eqref{grad}. This is the technique we shall apply to obtain
$A^{(4)}$, $A^{(5)}$ later in this paper. We should note that $A$ has no
explicit two-loop contributions; however there is a one-loop (free-field) contribution\cite{Sternew,Stertwo}
\footnote{See Refs.\cite{Toms,JacKod} for earlier perturbative calculations on six-dimensional curved spacetime.}.
At this point we have all the information required for our computation of $A^{(4)}$, but we shall postpone this to Section 4 where we shall explain the general method we shall use to compute both $A^{(4)}$ and $A^{(5)}$.

\section{Three-loop results}
In this section we shall give our explicit results for the three-loop $\beta$-functions for the theory Eq.~\eqref{lag}, computed using $\MSbar$. We start by describing the computational methods used\footnote{Similar methods have recently been applied to the computation of the four-loop $\MSbar$ $\beta$-function for this theory\cite{johnnew}.}, and we shall also 
take the opportunity to describe the MOM scheme (which will feature in later sections) and how to adapt our methods to obtain the MOM $\beta$-functions. The reader uninterested in the technical details of the Feynman diagram calculations may skip to the paragraph containing Eq.~\eqref{BF}.

 We base our calculations on the original work of Refs.~\cite{jag1,jag2}, although our
couplings are more general than those articles. However, one can always make
contact with the results of Refs.~\cite{jag1,jag2} by setting
$g^{ijk}$~$=$~$g d^{ijk}$ where $g$ was the coupling constant and $d^{ijk}$ was
a group valued object which is completely symmetric in its indices like
$g^{ijk}$. For example, if $\phi^i$ took values in the adjoint representation
of $SU(N_{\! c})$ then $d^{ijk}$ would be the corresponding totally symmetric
rank $3$ colour tensor of that group. From the point of view, however, of
constructing the RG functions it is the evaluation of the
underlying Feynman integrals which is required. For this aspect it is
appropriate to focus for the moment on the basic $\phi^3$ Lagrangian
\begin{equation}
L ~=~ \frac{1}{2} \left( \partial_\mu \phi \right)^2 ~+~ \frac{g}{6} \phi^3 ~.
\label{lagphi3}
\end{equation}
Previously this was considered in Refs.~\cite{jag1,jag2,jag3} to three loops and the
wave function and coupling constant renormalisation constants were deduced from
the divergences in the $2$- and $3$-point functions. For the latter the pole
structure was determined by exploiting specific properties of the six
dimensional spacetime. Briefly it was possible to nullify the external momentum
of one of the legs of the $3$-point functions to reduce the evaluation of the
graph to a $2$-point function. Such integrals are more straightforward to
determine through knowledge of the $2$-point evaluation. The point is that
ordinarily the nullification of an external momentum leads to infrared
problems. For instance, a propagator in a Feynman integral of the form
$1/(k^2)^2$ leads immediately to infrared singularities in the associated four-dimensional
integral where $k$ is the loop momentum. In six dimensions,
however, this is not the case. The corresponding propagator where such an
infrared issue would appear in that dimension is $1/(k^2)^3$. Therefore, the
nullification used in Ref.~\cite{jag2}, which is a simple application of the
infrared rearrangement technique, is perfectly valid in six dimensions for
Eq.~(\ref{lag}) and Eq.~(\ref{lagphi3}). Moreover, this method is sufficient to
determine the RG functions in the $\MSbar$ scheme.

However, as our focus here will not be restricted to $\MSbar$ but will include
the momentum subtraction (MOM) scheme we will have to determine the Feynman
integrals contributing to the vertex function for a non-nullified external
momentum and to the finite part in $\epsilon$ (where $d=6-2\epsilon$). Ahead of the description of the
computational tools we use it is appropriate to recall the definition of the
MOM scheme as this informs the integral evaluation. The particular MOM scheme
we use is that developed in Ref.~\cite{jag4} for Quantum Chromodynamics. However, we
note that it was used prior to that at three loops for Eq.~(\ref{lag}) in
Ref.~\cite{jag5} when the specific coupling tensor corresponded to an
$SU(3)$~$\times$~$SU(3)$ symmetry group. First, we recall that in the minimal
subtraction scheme the renormalisation constants are defined at a subtraction
point in such a way that only the poles with respect to the regularizing
parameter are included in the renormalisation constant. The $\MSbar$ scheme is a
variant on this where a specific finite part is also absorbed into the
renormalisation constants. This additional number, which is
$\ln(4\pi e^{-\gamma})$ where $\gamma$ is the Euler-Mascheroni constant, in
effect corresponds to a trivial rescaling of the coupling constant. The
renormalisation constants of the MOM scheme by contrast are defined at a
particular subtraction point in such a way that after renormalisation at that
point there are no $O(g)$ contributions to the Green's functions\cite{jag4}.
In other words the Green's functions are set to their tree values at the
subtraction point\cite{jag4}. The specific subtraction point used in Ref.~\cite{jag4} for
the $3$-point vertex functions is that where the squared momenta of all the
external legs are equal. Moreover, they are set equal to $(-\mu^2)$ where $\mu$
is the mass scale introduced to ensure the coupling constant or tensor is
dimensionless in general $d$ dimensions. To achieve this for Eq.~(\ref{lag}) requires
evaluating all the $2$-point function and vertex graphs to the finite parts in
$\epsilon$; the latter being a more involved computation than the former. While this is the
canonical definition of the MOM scheme we note
that one can have variations on it. For instance, it is acceptable to have a
scheme where the renormalisation constants associated with $2$-point functions
are defined in an $\MSbar$ way but the coupling-constant renormalisations are
determined using the MOM definition or vice versa. While it is possible to
study such hybrid schemes in order to consider different applications
of our general formalism, we will concentrate here purely on the $\MSbar$ and MOM schemes. However, we
will provide enough details in the evaluation of the symmetric-point $3$-point
integrals to allow an interested reader to explore these hybrid schemes
independently. In mentioning that a MOM scheme had been considered in
Ref.~\cite{jag5} we need to clarify this in light of our discussion. The
renormalisation of Ref.~\cite{jag5} was at three loops in a MOM scheme but for a
very specific coupling tensor. The consequence of the choice of the
$SU(3)$~$\times$~$SU(3)$ colour group is that there were no one or two loop
vertex graphs to be evaluated at the symmetric point. At three loops due to the
special symmetry properties there was only one graph to evaluate at the
symmetric point, and since it was primitively divergent the evaluation was straightforward\cite{jag5}.
Therefore, in considering the general
cubic theory in six dimensions Eq.~(\ref{lag}) we are filling in the gap in
the computation of the lower loop integrals for the momentum subtraction scheme
analysis.

One of our aims is the study of the $a$-theorem up to third order in various
RG schemes; it may appear a formidable task to actually {\em calculate} the MOM
$\beta$-function for Eq.~(\ref{lag}) at this order given the previous
discussion. However, it is possible to determine it purely from the {\em two}
loop renormalisation of the vertex functions in this scheme. To
illustrate the process, we consider for the moment
the simpler theory Eq.~(\ref{lagphi3}), postponing the general case
of Eq.~(\ref{lag}) to Sect. 5. After choosing a renormalisation scheme
for a given theory, any RG quantity will be a
function of the couplings in that scheme.
However, the expressions in different schemes must be related and this is
achieved by a conversion function. If we denote the coupling constant in one
scheme by $g$ and that in another by $\bar{g}$ then the conversion function
$C_g(g)$ is defined by
\begin{equation}
C_g(g) ~=~ \frac{\partial \bar{g}(g)}{\partial g}
\label{condef}
\end{equation}
where the bare coupling constant $g_{\mbox{\footnotesize{o}}}$ is related to
the two renormalized coupling constants by
\begin{equation}
g_{\mbox{\footnotesize{o}}} ~=~ \mu^\epsilon Z_g \, g ~=~
\mu^\epsilon \bar{Z}_g \, \bar{g} ~.
\end{equation}
$\bar{g}(g)$ has the form
\begin{equation}
\bar{g}(g) ~=~ g ~+~ \sum_{n=1}^\infty h_n g^{2n+1}
\label{mapexp}
\end{equation}
where the coefficients $h_n$ are related to the finite parts of the $Z_g$ and
$\bar{Z}_g$. Once the expansion has been established from the explicit
renormalisation then the respective $\beta$-functions are related by
\begin{equation}
\bar{\beta}(\bar{g}) ~=~ \left[ \beta(g) C_g(g)
\right]_{ g \rightarrow \bar{g} }
\label{betarel}
\end{equation}
where the mapping means that the coupling constant $g$ is mapped back to
$\bar{g}$ via the inverse of Eq.~(\ref{mapexp}). Now returning to our problem of
computing the MOM $\beta$-function, if $\bar{g}$ is the MOM
coupling constant and $g$ is that in the $\MSbar$ scheme, 
it is clear from Eq.~(\ref{betarel}) that only $h_1$ and $h_2$ are
required to find $\bar{\beta}(\bar{g})$ at three loops . These may be
derived from the finite part of
$\bar{Z}_g$ up to two loops which in turn derives from the finite parts of the one and two-loop 
vertex functions at the symmetric subtraction point. The coupling redefinition required 
to generate the three-loop MOM $\beta$-function will be presented in Sect. 5, and the three-loop
MOM $\beta$-function itself is given in Appendix A.

We now turn to the algorithm we used to evaluate the two and three loop Feynman
integrals to the requisite orders in $\epsilon$ to determine the
RG functions in $\MSbar$ and MOM. The method we use is to
apply the Laporta algorithm \cite{jag6} to the two and three loop graphs
contributing to the $2$-point and vertex functions. This method systematically
integrates by parts all the graphs in such a way that they are algebraically
reduced to a basic set of what is termed master integrals. Then the $\epsilon$-expansions
of the latter are substituted to complete the computation. The
masters have to be determined by direct methods or one which does not use
integration by parts and this is the more demanding aspect of the calculation. The
version of the Laporta algorithm which we used was {\sc Reduze}
\cite{jag7,jag8}. To handle the surrounding tedious algebra we used the
symbolic manipulation language {\sc Form} \cite{jag9,jag10}. The whole
evaluation proceeded automatically by generating all the Feynman diagrams
electronically with the {\sc Qgraf} package \cite{jag11}. While we have
summarised what is now a standard procedure to carry out multiloop Feynman
graph evaluation, the novel feature here is finding a method to access the six
dimensional master integrals. The most straightforward way to proceed is to
realise that if the problem was in four dimensions then the corresponding
masters are already known. For instance, the three loop $2$-point functions
were developed for the {\sc Mincer} package \cite{jag12} which was encoded in
{\sc Form} in Ref.~\cite{jag13}. Equally the two loop $3$-point function masters
were determined over a period of years in Refs.~\cite{jag14,jag15,jag16,jag17,jag18}.
While the algorithms which were developed to determine such four-dimensional
masters could in principle be extended to six dimensions, in practice this
would be tedious. Instead we have used the Tarasov method developed in
Refs.~\cite{jag19,jag20}. This allows one to relate a scalar (master) integral in
$d$-dimensions to integrals with the same propagator topology in $(d+2)$
dimensions. The caveat is that the integrals in the higher-dimensional case
have increased powers of their propagators. However, the Laporta algorithm
\cite{jag6} can be applied to them in order to reduce them to the
corresponding master in that dimension and integrals which involve masters
where the number of propagators in the topology have been reduced. In summary a
four dimensional master can be related to its unknown six-dimensional cousin
plus already determined lower-level six-dimensional masters. Hence one can
algebraically solve the system for the masters to the order in $\epsilon$ which
they are required for the $2$-point and vertex functions. One feature which
invariably arises in the use of integration by parts is the appearance of
spurious poles. Consequently in the determination of the masters in six
dimensions as well as the reduction of the Feynman graphs contributing to the
RG functions, one sometimes has to evaluate the masters
{\em beyond} the finite part in $\epsilon$. In the appendix we record the
expressions for the nontrivial one and two loop $3$-point masters; though we
stress that the expressions given there are for the pure integral. There has
been no subtraction of subgraphs as was the case in the results presented for
graphs in Refs.~\cite{jag2,kleb}. This is primarily because in the automatic
{\sc Form} programmes to determine the RG functions we use
the algorithm of Ref.~\cite{jag22} to implement the renormalisation automatically in
our two schemes. Briefly this is achieved by performing the calculation in
terms of the bare coupling constant or tensor. Then what would be termed
counterterms are introduced automatically by rescaling the bare parameter to
the renormalized one. The so-called constant of proportionality would
ordinarily correspond to the coupling constant renormalisation constant which
therefore appends the necessary counterterms. Finally, the definition of the
renormalisation constant is implemented at this last stage. At whatever loop
order one is working to, the remaining undetermined counterterm is defined
according to whether the scheme is $\MSbar$ or MOM.

We can now present our three-loop results for $\MSbar$.  At this order we require several new tensors. Those contributing to the one-particle irreducible (1PI) terms in the $\beta$-function are
\begin{align}
g_{(3e)}^{ijk} ~=~ g_{(2b)}^{ilm}g_{22}^{jlkm} ,~~~ g_{(3f)}^{ijk} &=g_{(2b)}^{lmi}g_{22}^{jlkm}  ,~~~ g_{(3g)}^{ijk} ~=~ g^{ipq} g_{(1a)}^{jpr} g_{(1a)}^{kqr} ,\nn
 g_{(3h)}^{ijk} ~=~ g_{(1a)}^{ilm}g_{(1A)}^{nq}g^{jln}g^{kmq},~~~ g_{(3i)}^{ijk} &= g_{(1A)}^{pq}  g_{(1a)}^{ipr}g_{22}^{jqkr} ,~~~ g_{(3j)}^{ijk} ~=~ g^{ilm}g^{jln}g^{kmq}g_{(2B)}^{nq} ,\nn
 g_{(3k)}^{ijk} ~=~g^{mil}_{(2c)}g_{22}^{jlkm} ,~~~ g_{(3l)}^{ijk} &=g^{ilm}_{(2c)}g_{22}^{jlkm}  ,~~~ g_{(3m)}^{ijk} ~=~ g^{iln}g^{jmq}g^{klm}g_{(2C)}^{nq} ,\nn
 g_{(3n)}^{ijk} ~=~ g^{iln}g^{jmq}g^{klm}g_{(2D)}^{nq} ,~~~ g_{(3o)}^{ijk} &= g_{(1A)}^{pq}g_{(1A)}^{rs} g_{22}^{jpkr} g^{iqs} ,~~~ g_{(3p)}^{ijk} ~=~ g^{ist}g_{(1A)}^{qt}g_{22}^{jrps}g_{22}^{kpqr}   ,\nn
  g_{(3q)}^{ijk} ~=~ g_{(1a)}^{irs} g_{22}^{jqps}g_{22}^{kpqr}  ,~~~ g_{(3r)}^{ijk} &= g^{ipq} g^{jrs} g_{22}^{knps} g_{(1a)}^{rnq} ,~~~ g_{(3s)}^{ijk} ~=~ g_{(2d)}^{ilm}g_{22}^{jlkm} ,\nn
   g_{(3t)}^{ijk} ~=~ g^{ipq} g_{22}^{jrps} g_{22}^{knrl} g_{22}^{nqls} ,~~~ g_{(3u)}^{ijk} &= g^{kpq} g_{22}^{jrsn} g_{22}^{ispl} g_{22}^{rlqn},
   \label{BF}
\end{align}
where
\be
g^{ijkl}_{22}=g^{ijm}g^{klm},\quad  g_{(2D)}^{ij}=g_{(1A)}^{im}g_{(1A)}^{mj},
\ee
and $g_{(1a)}$,   $g_{(1A)}$ etc are defined in Eqs.~\eqref{tensa}, \eqref{tensc}. The tensors contributing to the anomalous dimension are
\begin{align}
 g_{(3D)}^{ij} ~=~ g^{ipq} g_{(2b)}^{qjp} ,~~~ g_{(3E)}^{ij} ~&=~ g^{ipq} g_{(2b)}^{jpq} ,~~ g_{(3F)}^{ij} ~=~ g^{ipq} g_{(2c)}^{jpq} ,\nn
 g_{(3G)}^{ij} ~=~ g^{imn}g_{(1A)}^{mp}g_{(1a)}^{pnj} ,~~~
 g_{(3G')}^{ij} ~&=~ g_{(1a)}^{inp}g_{(1A)}^{mp}g^{jmn} ,\quad
  g_{(3H)}^{ij} ~=~ g_{22}^{ipjq} g_{(2B)}^{pq} ,\nn
  g_{(3I)}^{ij} ~=~ g^{ipq} g_{(2d)}^{jpq} ,~~ g_{(3J)}^{ij} ~&=~ g^{imn}g_{(1A)}^{mp}g_{(1A)}^{nq}g^{jpq} ,~~~
  ~~~ g_{(3K)}^{ij} ~=~ g_{22}^{ipjq} g_{(2C)}^{pq} ,\nn
   g_{(3L)}^{ij} ~&=~g_{22}^{injq}g_{(2D)}^{nq} .~~~
 \label{WF}
 \end{align}
 The notation for the two-index tensors in Eq.~\eqref{WF} matches the diagrams in Figure 7 of Ref.~\cite{kleb}, so that $g_{(3D)}^{ij}$ corresponds to the tensor structure of Fig.~7(d), and so on. The notation for the three-index tensors $g^{ijk}_{(3e)}\ldots g^{ijk}_{(3u)}$ in Eq.~\eqref{BF} similarly matches the diagrams in Figures 8 and 9 of Ref.~\cite{kleb}; and furthermore, the indices $i,j,k$ are arranged to run anti-clockwise around the diagram, with $i$ at the top. Finally, $g^{ijk}_{(3D)}\ldots g^{ijk}_{(3L)}$, as defined in Eq.~\eqref{anomdim} in terms of 
$g^{ij}_{(3D)}\ldots g^{ij}_{(3L)}$, correspond to three-point diagrams with an insertion of the corresponding wave-function renormalisation diagram of Fig.~7 of Ref.~\cite{kleb}.

As in Section 2, we write the three-loop
$\beta$-function as
\be
\beta^{(3)}=c_{(3e)}g_{(3e)}+\ldots+c_{(3u)}g_{(3u)}+c_{(3D)}g_{(3D)}+\ldots+c_{(3L)}g_{(3L)}.
\label{betathree}
\ee
The  coefficients  in Eq.~\eqref{betathree} corresponding to anomalous dimension contributions are given by
\begin{align}
c_{(3D)} ~=~ ~ \tfrac{7}{864},\quad
c_{(3E)} = ~ \tfrac{71}{1728},\quad&
c_{(3F)} ~=~- \tfrac{103}{10368},\quad
c_{(3G)} ~=~ c_{(3G')}~=~ -\tfrac{1}{108},\nn
c_{(3H)} =-\tfrac{121}{5184},\quad
c_{(3I)} = ~ \tfrac{7}{96} - \tfrac{1}{24} \zeta(3),\quad&
c_{(3J)} ~=~  \tfrac{23}{62208},\quad
c_{(3K)} ~=~ \tfrac{103}{7776},\nn
c_{(3L)} = -\tfrac{13}{31104},\quad
 \label{betthree}
 \end{align}
where $\zeta(z)$ is the Riemann $\zeta$-function, and the remaining coefficients are given by
\begin{align} 
c_{(3e)} ~=&~ -\tfrac{3}{8},\quad
c_{(3f)} ~=~  \tfrac{1}{4},\nn
c_{(3g)} ~=~ \tfrac{5}{16},\quad
c_{(3h)} ~=~ -\tfrac{47}{864},\quad&
c_{(3i)} ~=~ -\tfrac{47}{432},\quad
c_{(3j)} ~=~  \tfrac{23}{288},\nn
c_{(3k)} ~=~  \tfrac{5}{27},\quad
c_{(3l)} =  \tfrac{11}{216},\quad&
c_{(3m)} ~=~ -\tfrac{19}{324},\quad
c_{(3n)} ~=~  \tfrac{11}{1728},\nn
c_{(3o)} =  \tfrac{11}{1728},\quad
c_{(3p)} ~=~ \tfrac{11}{144},\quad&
c_{(3q)} ~=~ -\tfrac{1}{16},\quad
c_{(3r)} ~=~ -\tfrac{23}{24} + \zeta(3),\nn
c_{(3s)} = -\tfrac{29}{48} + \tfrac{1}{2} \zeta(3),\quad
c_{(3t)} ~=~ -1,\quad&
c_{(3u)} ~=~  \tfrac{1}{3} - \zeta(3)  .
 \label{betthreea}
 \end{align}
We have computed all the coefficients in Eqs.~\eqref{betthree},  \eqref{betthreea} explicitly and independently; and we have checked that we reproduce the wave function and $\beta$-function results of Refs. \cite{kleb} and
\cite{jag1,jag2} (in the latter case, after we specialize to the corresponding restriction on the group theory structure used
there). Although in Ref.~\cite{kleb} the final $\beta$-function results are given for two particular theories, the general results can be constructed from the individual diagrammatic results. This is largely the case for Ref.~\cite{jag1,jag2} too; however, the results for the pairs $\{c_{(3e)}, c_{(3f)}\}$, $\{c_{(3k)}, c_{(3l)}\}$, and  $\{c_{(3D)}, c_{(3E)}\}$, are presented together and cannot be separated.

\section{The $a$-function beyond leading order}
We now turn to the derivation of $A^{(4)}$, $A^{(5)}$.
At two and three loops, we have avoided any diagrammatic computations using the methods of Ref.~\cite{Analog,Analoga},
as explored in the six-dimensional context in Refs.~\cite{Sternew, Stertwo}; instead we have proceeded to infer $A$ by imposing Eq.~\eqref{grad}. Beyond leading order (corresponding to Eq.~\eqref{Athree}) we need to take into account potential higher order corrections to $T_{IJ}$.  We suppress the details here but the calculation proceeds along similar lines to those presented in full in the four-dimensional case in Refs.\cite{OsbJacnew, JacPoole}; a similar method was used in the pioneering work of Wallace and Zia\cite{Wallace}. At next-to-leading order the general form of the $a$-function is given by

\be
A^{(4)}=\lambda\left[-\tfrac{1}{12}A^{(4)}_1+a_2A^{(4)}_2
+a_3A^{(4)}_3+a_4A^{(4)}_4
+a_5A^{(4)}_5+\alpha_1\beta^{(1)ijk}\beta^{(1)ijk}\right],
\label{Afour}
\ee

\begin{table}[h]
	\setlength{\extrarowheight}{0.5cm}
	\setlength{\tabcolsep}{24pt}
	\hspace*{-8.25cm}
	\centering
	\resizebox{6.7cm}{!}{
		\begin{tabular*}{20cm}{cccccc}
			\begin{picture}(162,162) (287,-207)
			\SetWidth{1.0}
			\SetColor{Black}
			\Line(306,-176)(357,-134)
			\Line(431,-175)(379,-134)
			\Arc(368,-126)(80,127,487)
			\Line(304,-78)(357,-120)
			\Line(430,-76)(379,-120)
			\Line(368,-47)(368,-205)
			\end{picture}
			&
			\begin{picture}(162,162) (287,-207)
			\SetWidth{1.0}
			\SetColor{Black}
			\Arc(368,-126)(80,127,487)
			\Arc(368,-127)(31.064,123,483)
			\Line(368,-46)(368,-96)
			\Line(345,-148)(304,-174)
			\Line(391,-148)(431,-174)
			\end{picture}
			&
			\begin{picture}(162,162) (287,-207)
			\SetWidth{1.0}
			\SetColor{Black}
			\Arc(368,-126)(80,127,487)
			\Arc[clock](269.598,-126.405)(82.403,51.407,-50.512)
			\Arc(466.565,-126)(81.565,128.311,231.689)
			\Line(289,-126)(352,-126)
			\end{picture}
			&
			\begin{picture}(162,162) (287,-207)
			\SetWidth{1.0}
			\SetColor{Black}
			\Arc(368,-126)(80,127,487)
			\Arc[clock](269.598,-126.405)(82.403,51.407,-50.512)
			\Arc(466.565,-126)(81.565,128.311,231.689)
			\Line(368,-46)(368,-206)
			\end{picture}
			&
			\begin{picture}(162,162) (287,-207)
			\SetWidth{1.0}
			\SetColor{Black}
			\Arc(368,-126)(80,127,487)
			\Arc[clock](367.5,-190.522)(43.525,-178.054,-361.946)
			\Arc(317.004,-93.238)(40.144,-134.235,68.065)
			\Arc[clock](421.537,-93.476)(39.647,-48.13,-246.928)
			\end{picture}
			&
			\begin{picture}(162,191) (287,-192)
			\SetWidth{1.0}
			\SetColor{Black}
			\Arc(368,-97)(80,127,487)
			\Line(368,-18)(368,-176)
			\CTri(352,-18)(368,-2)(384,-18){Black}{Black}\CTri(352,-18)(368,-34)(384,-18){Black}{Black}
			\CTri(352,-175)(368,-159)(384,-175){Black}{Black}\CTri(352,-175)(368,-191)(384,-175){Black}{Black}
			\Text(394,-3)[]{\Huge{\Black{$(1)$}}}
			\Text(395,-193)[]{\Huge{\Black{$(1)$}}}
			\end{picture}
			\\
			&
			&
			&
			&
			&
			\\
			{\Huge $A^{(4)}_{1}$}
			&
			{\Huge $A^{(4)}_{2}$}
			&
			{\Huge $A^{(4)}_{3}$}
			&	
			{\Huge $A^{(4)}_{4}$}
			&	
			{\Huge $A^{(4)}_{5}$}
			&	
			{\Huge $A^{(4)}_{6}$}
		\end{tabular*}
	}
	\caption{Contributions to $A^{(4)}$}
	\label{A4pics}	
\end{table}

\begin{table}[h]
	\setlength{\extrarowheight}{0.5cm}
	\setlength{\tabcolsep}{24pt}
	\hspace*{-3.5cm}
	\centering
	\resizebox{6.7cm}{!}{
		\begin{tabular*}{20cm}{cccc}
			\begin{picture}(162,179) (287,-190)
			\SetWidth{1.0}
			\SetColor{Black}
			\Arc(368,-109)(80,127,487)
			\Line(368,-109)(368,-29)
			\Line(368,-109)(304,-157)
			\Line(368,-109)(432,-157)
			\SetWidth{2.0}
			\Line(352,-13)(384,-45)
			\Line(384,-13)(352,-45)
			\SetWidth{1.0}
			\CTri(345.373,-109)(368,-86.373)(390.627,-109){Black}{Black}\CTri(345.373,-109)(368,-131.627)(390.627,-109){Black}{Black}
			\Text(368,-150)[]{\Huge{\Black{$(1)$}}}
			\end{picture}
			&
			\begin{picture}(182,170) (267,-205)
			\SetWidth{1.0}
			\SetColor{Black}
			\Arc(368,-118)(80,127,487)
			\Arc[clock](269.598,-118.405)(82.403,51.407,-50.512)
			\Arc(466.565,-118)(81.565,128.311,231.689)
			\SetWidth{2.0}
			\Line(306,-37)(338,-69)
			\Line(338,-37)(306,-69)
			\SetWidth{1.0}
			\CTri(301.373,-181)(324,-158.373)(346.627,-181){Black}{Black}\CTri(301.373,-181)(324,-203.627)(346.627,-181){Black}{Black}
			\Text(282,-185)[]{\Huge{\Black{$(1)$}}}
			\end{picture}
			&
			\begin{picture}(187,171) (287,-205)
			\SetWidth{1.0}
			\SetColor{Black}
			\Arc(368,-117)(80,127,487)
			\Arc[clock](269.598,-117.405)(82.403,51.407,-50.512)
			\Arc(466.565,-117)(81.565,128.311,231.689)
			\SetWidth{2.0}
			\Line(306,-36)(338,-68)
			\Line(338,-36)(306,-68)
			\SetWidth{1.0}
			\CTri(396.373,-181)(419,-158.373)(441.627,-181){Black}{Black}\CTri(396.373,-181)(419,-203.627)(441.627,-181){Black}{Black}
			\Text(463,-182)[]{\Huge{\Black{$(1)$}}}
			\end{picture}
			&
			\begin{picture}(183,169) (287,-200)
			\SetWidth{1.0}
			\SetColor{Black}
			\Arc(368,-119)(80,127,487)
			\Arc[clock](269.598,-119.405)(82.403,51.407,-50.512)
			\Arc(466.565,-119)(81.565,128.311,231.689)
			\SetWidth{2.0}
			\Line(306,-38)(338,-70)
			\Line(338,-38)(306,-70)
			\SetWidth{1.0}
			\CTri(394.373,-55)(417,-32.373)(439.627,-55){Black}{Black}\CTri(394.373,-55)(417,-77.627)(439.627,-55){Black}{Black}
			\Text(463,-54)[]{\Huge{\Black{$(1)$}}}
			\end{picture}
			\\
			{\Huge $T^{(3)}_{1}$}
			&
			{\Huge $T^{(3)}_{2}$}
			&
			{\Huge $T^{(3)}_{3}$}
			&	
			{\Huge $T^{(3)}_{4}$}
		\end{tabular*}
	}
	\caption{Next-to-leading-order metric terms $T^{(3)}$}
	\label{Tmetric}	
\end{table}

\noindent where $\lambda$ is defined in Eq.~\eqref{Athreea}. The individual contributions to $\tilde{A}^{(4)}$, depicted above in Table \ref{A4pics}, are given by
\begin{align}
A^{(4)}_1=g^{ijk}g^{ijk}_{(2d)},\quad A^{(4)}_2&=g^{ijk}g^{kij}_{(2b)},\quad A^{(4)}_3=g^{ijk}g^{ijk}_{(2c)},\nn
A^{(4)}_4=g^{ijk}g^{ijk}_{(2C)},\quad A^{(4)}_5&=g_{(1A)}^{ij}g_{(1A)}^{jk}g_{(1A)}^{ki},
\end{align}
with the tensor structures again defined in Eqs.~\eqref{tensa}, \eqref{tensc}, \eqref{anomdim}, and (from Eq.~\eqref{betatwo})
\be
\beta^{(1)ijk}\beta^{(1)ijk}=A^{(4)}_2-\tfrac12A^{(4)}_3+\tfrac{1}{24}A^{(4)}_4+\tfrac{1}{48}A^{(4)}_5.
\ee

Correspondingly the tensor $T_{IJ}$ in Eq.~\eqref{grad} is automatically symmetric at this order (so that
$G_{IJ}=T_{IJ}$) and may be written in the form
\be
T_{ijk,lmn}^{(3)}=\sum_{\alpha=1}^{4}t^{(3)}_{\alpha}\left(T_{\alpha}^{(3)}\right)_{ ijk,lmn},
\label{metthree}
\ee
where the individual structures which may arise are depicted in
Table~\ref{Tmetric}. Here the diagrams represent $t^{(3)}_{\alpha} \small( T_{\alpha}^{(3)} \small)_{ijk,lmn}\beta^{ijk}(dg)^{lmn}$
for $\alpha=1\ldots4$. A cross denotes $(dg)^{ijk}$ and a diamond represents $\beta^{ijk}$.

A careful analysis leads to a system of linear equations whose solution imposes a single consistency condition on the $\beta$-function coefficients:
\be 6c_{(2C)}+c_{(2c)}+c_{(2B)}=0.
\label{constwo}
\ee
 This is satisfied by the $\MSbar$ coefficients as given by Eq.~\eqref{cmstwo}. Similar integrability conditions (on three-loop $\beta$-function coefficients) were found in Ref.~\cite{OsbJacnew}.
In Eq.~\eqref{Afour}, $\alpha_1$ is arbitrary, reflecting the general freedom expressed in Eq.~\eqref{free};  and in particular for $\MSbar$ we have
\be
a_2=-\tfrac18+\alpha_1,\quad a_3=\tfrac{7}{48}-\tfrac12\alpha_1, \quad a_4=-\tfrac{7}{288}+\tfrac{1}{24}\alpha_1,\quad a_5=-\tfrac{5}{864}+\tfrac{1}{48}\alpha_1.
\ee
 Our methods therefore specify $A$ up to the freedom expressed in Eq.~\eqref{free}.
The metric coefficients in Eq.~\eqref{metthree} are given by
\be
t^{(3)}_2=-\tfrac{7}{24}\lambda+\tfrac12\alpha_1,\quad
t^{(3)}_3+t^{(3)}_4=-\tfrac{1}{8}\lambda+\alpha_1,\quad t^{(3)}_1=-6\alpha_1,
\label{tcoeffs}
\ee

where once more $\lambda$ is defined in Eq.~\eqref{Athreea}. The metric coefficients therefore also reflect the freedom expressed in Eq.~\eqref{free}; but there is an additional arbitrariness since only the combination $t^{(3)}_3+t^{(3)}_4$
is determined.

At the next order (corresponding to the three-loop $\beta$-function) we need to face the possibility that $\beta^I$ in Eq.~\eqref{grad} should be replaced by a generalisation $B^I$ due to the invariance of the Lagrangian Eq.~\eqref{lag} under $O(N)$ transformations of the real fields $\phi^i$. It was shown in Ref.~\cite{Analoga} that in this situation, in the general case with couplings $g^I$ we have
\be
\beta^I\rightarrow B^I=\beta^I-(vg)^I
\ee
where $v$ is an element of the Lie algebra of the symmetry group. In the case at hand, this corresponds to
\be
\beta^{ijk}\rightarrow B^{ijk}=\beta^{ijk}-v^{l(i}g^{jk)l}
\label{Bdef}
\ee
where $v$ is an antisymmetric matrix. In principle $v$ could be computed using similar methods to those described in Ref.~\cite{Analoga} and carried out explicitly in the four-dimensional case in Ref.~\cite{OsbJacnew}; but it is clear {\it a priori} that the relevant tensor structures are the same as those appearing in the three-loop anomalous dimension. Since most of those are symmetric, the only possible 1PI contributions to $v$ correspond to
\be
v=c^v_{(3G)}(g_{(3G)}-g_{(3G')})
\label{vdef}
\ee
with $g_{(3G)}$, $g_{(3G')}$ as defined in Eq.~\eqref{WF}.

The solution of Eq.~\eqref{grad} leads to a complex system of linear equations whose solution imposes several consistency conditions on the $\beta$-function coefficients:

\begin{align}
c_{(3q)}-c_{(3s)}-12c_{(3I)}&=-6c_{(2B)},\nn
c_{(3r)}-2c_{(3s)}+12c_{(3p)}&=12c_{(2c)},\nn
c_{(3e)}-c_{(3g)}-24c_{(3h)}-144c_{(3o)}-72Z&=-
3(c_{(2B)}+2c_{(2c)}),\nn
c_{(3e)}-c_{(3g)}-6c_{(3i)}+6c_{(3k)}+72Z&=3c_{(2B)}+144c_{(2c)}^2,\nn
c_{(3j)}+12c_{(3n)}-c_{(3E)}+12c_{(3H)}+36c_{(3K)}+72c_{(3L)}&\nn-12(c_{(3\rho)}+c_{(3\sigma)})
-72(c_{(3\tau)}+c_{(3\chi)})
+6Z&=12(2c_{(2c)}^2+2c_{(2c)}c_{(2B)}-c_{(2B)}^2),\nn
2c_{(3h)}+6c_{(3m)}-12c_{(3n)}+18c_{(3o)}+c_{(3D)}+12c_{(3F)}+72c_{(3J)}&\nn+36c_{(3K)}-72c_{(3L)}
&=\tfrac{11}{144}[1+24(c_{(2B)}-c_{(2c)})],\nn
c_{(3e)}-\tfrac12c_{(3f)}+6c_{(3k)}-12c_{(3l)}&=0,\nn
c_{(3j)}+6c_{(3m)}+6c_{(3H)}+36c_{(3K)}&=12c_{(2B)}c_{(2c)},\nn
c_{(3h)}-c_{(3i)}+c_{(3l)}-c_{(3D)}-12c_{(3F)}+12Z&=12c_{(2c)}(c_{(2B)}+2c_{(2c)}),
\label{conthree}
\end{align}
and
\be
12c^v_{(3G)}+6(c_{(3G)}-c_{(3G')})+12(c_{(3\rho)}-c_{(3\sigma)})+72(c_{(3\tau)}-c_{(3\chi)})=c_{(3j)}+6c_{(3m)}+12c^{2}_{(2c)},
\label{conthreea}
\ee
where
\be
Z=c_{(3G)}+c_{(3G')}-c_{(3o)}+\tfrac16c_{(3E)}-2c_{(3F)}+12c_{(3J)}.
\ee
We have included in our calculations potential contributions to the three-loop $\beta$-functions defined according to Eq.~\eqref{anomdim} in terms of one-particle-reducible (1PR) anomalous dimension
structures given by
\be
 g_{(3\rho)}^{ij} ~=~ g_{(2B)}^{il}g_{(1A)}^{lj} ,\quad  g_{(3\sigma)}^{ij} ~=~ g_{(1A)}^{il}g_{(2B)}^{lj} ,\quad
 g_{(3\tau)}^{ij}=g_{(2C)}^{ik}g_{(1A)}^{kj},\quad  g_{(3\chi)}^{ij}=g_{(1A)}^{ik}g_{(2C)}^{kj},
 \ee
 using again tensors defined earlier in Eq.~\eqref{tensc}. Such contributions cannot of course arise in $\MSbar$ but are potentially present in other schemes, in particular the MOM scheme which we shall be considering later as an example. (There are other potential 1PR contributions in a general scheme, but we have included only the ones which are relevant for MOM.) In deriving Eqs.~\eqref{conthree},  \eqref{conthreea}, we have imposed the two-loop consistency condition Eq.~\eqref{constwo} and also used the $\MSbar$ values of Eq.~\eqref{cmstwo} for all two-loop $\beta$ function coefficients except $c_{(2B)}$ and $c_{(2c)}$, since as we shall see later (in Eq.~\eqref{redone}), these are the only scheme-dependent values.
 The conditions in Eq.~\eqref{conthree} are readily checked to be satisfied by the $\MSbar$ three-loop $\beta$-function coefficients given in Eq.~\eqref{betthree}, \eqref{betthreea};  however, Eq.~\eqref{conthreea} may only be satisfied within $\MSbar$ by taking a non-zero value of $c_{(3G)}^v$, namely
\be
 \left( c_{(3G)}^{v} \right)_{\scriptsize{\MSbar}}=-\tfrac{137}{10368}.
\label{vval}
\ee
We shall assume this value from now on, though of course it would be reassuring to compute it directly along the lines of Ref.~\cite{FortinC}.

Given that the consistency conditions are satisfied, we may solve Eq.\eqref{grad} for the $a$-function.
With the three-loop $\MSbar$ coefficients we find
\be
A^{(5)}=\sum_{i=1}^{16}a^{(5)}_iA_{i}^{(5)},
\label{athree}
\ee
where
\begin{align}
A^{(5)}_1&=g^{ijk}g_{(3I)}^{ijk}=g^{ijk}g_{(3p)}^{ijk},\quad&
A^{(5)}_9&=g^{ijk}g_{(3M)}^{ijk}=g^{ijk}g_{(3N)}^{ijk}=g^{ijk}g_{(3n)}^{ijk}, \nn
A^{(5)}_2&=g^{ijk}g_{(3q)}^{ijk}=g^{ijk}g_{(3r)}^{ijk}=g^{ijk}g_{(3s)}^{ijk}, \quad& A^{(5)}_{10}&=g^{ijk}g_{(3j)}^{ijk}, \nn
A^{(5)}_3&=g^{ijk}g_{(3u)}^{ijk},\quad& A^{(5)}_{11}&=g^{ijk}g_{(3H)}^{ijk}=g^{ijk}g_{(3m)}^{ijk}, \nn
A^{(5)}_4&=g^{ij}_{(1A)}g^{jk}_{(1A)}g^{kl}_{(1A)}g^{li}_{(1A)}, \quad& A^{(5)}_{12}&=g^{ijk}g_{(3e)}^{ijk}=g^{ijk}g_{(3f)}^{ijk}=g^{ijk}g_{(3g)}^{ijk}, \nn
A^{(5)}_5&=g^{ijk}g_{(3L)}^{ijk}=g^{ijk}g^{ijk}_{(3P)}=g^{ijk}g_{(3Q)}^{ijk},\quad& A^{(5)}_{13}&=g^{ijk}g_{(3D)}^{ijk}=g^{ijk}g_{(3h)}^{ijk}=g^{ijk}g_{(3k)}^{ijk}=g^{ijk}g_{(3l)}^{ijk}, \nn
A^{(5)}_6&=g^{ijk}g_{(3K)}^{ijk}, \quad& A^{(5)}_{14}&=g^{ijk}g_{(3E)}^{ijk}=g^{ijk}g_{(3i)}^{ijk}, \nn
A^{(5)}_7&=g^{ijk}g_{(3J)}^{ijk},\quad& A^{(5)}_{15}&=g^{ijk}g_{(3o)}^{ijk}=g^{ijk}g_{(3G)}^{ijk}, \nn
A^{(5)}_8&=g^{ijk}g_{(3t)}^{ijk}, \quad& A^{(5)}_{16}&=g^{ijk}g_{(3F)}^{ijk},
\end{align}
with the tensor structures $g^{ijk}_{(3p)}$ etc as defined in Eq.~\eqref{BF}. The invariants $A^{(5)}_1\ldots
A^{(5)}_{16}$ are depicted in Table~\ref{fig2}. The coefficients $a^{(5)}_i$ in Eq.~\eqref{athree} are given by

\begin{table}[h]
	\setlength{\extrarowheight}{0.5cm}
	\setlength{\tabcolsep}{24pt}
	\hspace*{-8.5cm}
	\centering
	\resizebox{7cm}{!}{
		\begin{tabular*}{20cm}{cccccc}
			\begin{picture}(162,162) (287,-207)
			\SetWidth{1.0}
			\SetColor{Black}
			\Line(306,-176)(347,-144)
			\Line(431,-175)(390,-143)
			\Arc(368,-126)(80,127,487)
			\Arc(368,-126)(21.024,155,515)
			\Line(368,-105)(368,-47)
			\Line(368,-205)(368,-147)
			\Line(304,-78)(346,-110)
			\Line(430,-76)(390,-108)
			\end{picture}
			&
			\begin{picture}(162,162) (287,-207)
			\SetWidth{1.0}
			\SetColor{Black}
			\Line(431,-175)(376,-133)
			\Arc(368,-126)(80,127,487)
			\Line(368,-178)(368,-47)
			\Line(304,-78)(360,-121)
			\Line(430,-76)(376,-121)
			\Arc[clock](368,-236)(58,133.603,46.397)
			\Line(306,-177)(360,-134)
			\end{picture}
			&
			\begin{picture}(162,162) (287,-207)
			\SetWidth{1.0}
			\SetColor{Black}
			\Line(431,-175)(376,-133)
			\Arc(368,-126)(80,127,487)
			\Line(368,-205)(368,-127)
			\Line(304,-78)(360,-121)
			\Line(430,-76)(376,-121)
			\Line(306,-177)(360,-134)
			\Line(288,-126)(448,-126)
			\end{picture}
			&
			\begin{picture}(162,162) (287,-207)
			\SetWidth{1.0}
			\SetColor{Black}
			\Arc(368,-126)(80,127,487)
			\Arc(317.173,-75.173)(45.767,-126.421,36.421)
			\Arc[clock](419.813,-76.426)(44.174,-53.642,-220.053)
			\Arc[clock](317.104,-177.896)(45.197,125.279,-35.279)
			\Arc(420.856,-176.125)(44.014,55.161,217.634)
			\end{picture}
			&
			\begin{picture}(162,162) (287,-207)
			\SetWidth{1.0}
			\SetColor{Black}
			\Arc(368,-126)(80,127,487)
			\Arc(317.173,-75.173)(45.767,-126.421,36.421)
			\Arc[clock](419.813,-76.426)(44.174,-53.642,-220.053)
			\Arc[clock](367.5,-225.5)(75.515,141.992,38.008)
			\Arc[clock](367,-231.273)(63.304,139.309,40.691)
			\end{picture}
			&
			\begin{picture}(162,162) (287,-207)
			\SetWidth{1.0}
			\SetColor{Black}
			\Arc(368,-126)(80,127,487)
			\Line(353,-48)(353,-205)
			\Line(384,-48)(384,-205)
			\Arc[clock](279.822,-127.493)(50.181,65.037,-62.491)
			\Arc(458.02,-127.534)(51.022,116.819,240.635)
			\end{picture}
			\\
			{\Huge $A^{(5)}_{1}$}
			&
			{\Huge $A^{(5)}_{2}$}
			&
			{\Huge $A^{(5)}_{3}$}
			&	
			{\Huge $A^{(5)}_{4}$}
			&
			{\Huge $A^{(5)}_{5}$}
			&
			{\Huge $A^{(5)}_{6}$}
			\\
			&
			&
			&
			&
			&
			\\
			\begin{picture}(162,162) (287,-207)
			\SetWidth{1.0}
			\SetColor{Black}
			\Arc(368,-126)(80,127,487)
			\Arc(368,-126)(19.209,141,501)
			\Line(368,-47)(368,-107)
			\Line(368,-145)(368,-205)
			\Arc[clock](259.746,-126.462)(76.255,47.768,-46.745)
			\Arc(480.652,-126)(79.658,135.332,224.668)
			\end{picture}
			&
			\begin{picture}(162,162) (287,-207)
			\SetWidth{1.0}
			\SetColor{Black}
			\Arc(368,-126)(80,127,487)
			\Arc(368,-126)(31.145,138,498)
			\Line(368,-46)(368,-95)
			\Line(336,-126)(288,-126)
			\Line(399,-126)(448,-126)
			\Line(368,-157)(368,-206)
			\end{picture}
			&
			\begin{picture}(162,162) (287,-207)
			\SetWidth{1.0}
			\SetColor{Black}
			\Arc(368,-126)(80,127,487)
			\Arc(317.714,-75.714)(45.659,-127.372,37.372)
			\Arc[clock](419.855,-76.386)(44.181,-53.716,-219.979)
			\Arc[clock](367.5,-225.534)(75.536,141.971,38.029)
			\Line(368,-150)(368,-206)
			\end{picture}
			&
			\begin{picture}(162,162) (287,-207)
			\SetWidth{1.0}
			\SetColor{Black}
			\Arc(368,-126)(80,127,487)
			\Arc[clock](269.598,-126.405)(82.403,51.407,-50.512)
			\Arc(466.565,-126)(81.565,128.311,231.689)
			\Line(289,-126)(352,-126)
			\Line(386,-126)(448,-126)
			\end{picture}
			&
			\begin{picture}(162,162) (287,-207)
			\SetWidth{1.0}
			\SetColor{Black}
			\Arc(368,-126)(80,127,487)
			\Arc[clock](269.598,-126.405)(82.403,51.407,-50.512)
			\Arc(466.565,-126)(81.565,128.311,231.689)
			\Line(289,-126)(352,-126)
			\Line(368,-46)(368,-206)
			\end{picture}
			&
			\begin{picture}(162,162) (287,-207)
			\SetWidth{1.0}
			\SetColor{Black}
			\Arc(368,-126)(80,127,487)
			\Arc(368,-127)(31.064,123,483)
			\Line(368,-46)(368,-96)
			\Line(345,-148)(304,-174)
			\Line(391,-148)(431,-174)
			\Line(337,-126)(399,-126)
			\end{picture}
			\\
			{\Huge $A^{(5)}_{7}$}
			&	
			{\Huge $A^{(5)}_{8}$}
			&
			{\Huge $A^{(5)}_{9}$}
			&
			{\Huge $A^{(5)}_{10}$}
			&
			{\Huge $A^{(5)}_{11}$}
			&	
			{\Huge $A^{(5)}_{12}$}
			\\
			&
			&
			&
			&
			&
			\\
			&
			\begin{picture}(162,162) (287,-207)
			\SetWidth{1.0}
			\SetColor{Black}
			\Arc(368,-126)(80,127,487)
			\Arc(368,-127)(31.064,123,483)
			\Line(368,-46)(368,-96)
			\Line(345,-148)(304,-174)
			\Line(391,-148)(431,-174)
			\Arc[clock](368,-217.114)(42.114,151.471,28.529)
			\end{picture}
			&
			\begin{picture}(162,162) (287,-207)
			\SetWidth{1.0}
			\SetColor{Black}
			\Arc(368,-126)(80,127,487)
			\Arc(368,-127)(31.064,123,483)
			\Line(345,-148)(304,-174)
			\Line(391,-148)(431,-174)
			\Arc(368,-71)(10.817,124,484)
			\Line(368,-46)(368,-61)
			\Line(368,-82)(368,-95)
			\end{picture}
			&
			\begin{picture}(162,162) (287,-207)
			\SetWidth{1.0}
			\SetColor{Black}
			\Line(410,-151)(438,-165)
			\Arc(368,-126)(80,127,487)
			\Line(368,-126)(368,-46)
			\Arc(336,-145)(11.402,142,502)
			\Arc(400,-145)(11.402,142,502)
			\Line(368,-127)(389,-139)
			\Line(368,-127)(346,-139)
			\Line(326,-151)(298,-165)
			\end{picture}
			&
			\begin{picture}(162,162) (287,-207)
			\SetWidth{1.0}
			\SetColor{Black}
			\Arc(368,-126)(80,127,487)
			\Arc(368,-127)(49.578,132,492)
			\Arc(368,-126)(19.723,120,480)
			\Line(368,-46)(368,-77)
			\Line(368,-177)(368,-206)
			\Line(319,-126)(348,-126)
			\Line(388,-126)(417,-126)
			\end{picture}
			&
			\\
			&
			{\Huge $A^{(5)}_{13}$}
			&
			{\Huge $A^{(5)}_{14}$}
			&
			{\Huge $A^{(5)}_{15}$}
			&	
			{\Huge $A^{(5)}_{16}$}
			&
		\end{tabular*}
	}
	\caption{Contributions to $A^{(5)}$}
	\label{fig2}	
\end{table}

\begin{align}
a^{(5)}_{1 }&= \left(\tfrac{9}{64}-\tfrac{1}{16}\zeta(3)\right)\lambda-\tfrac14\alpha_1, \quad& a^{(5)}_{9}&=\tfrac{47}{1152}\lambda+\tfrac{1}{36}\alpha_1+\tfrac{1}{144}\atil_1-\tfrac16\atil_2-\tfrac16\atil_3, \nn
a^{(5)}_{2 }&=\left( -\tfrac{29}{48}+\tfrac12\zeta(3)\right)\lambda+\alpha_1, \quad& a^{(5)}_{10 }&= -\tfrac{23}{576}\lambda-\tfrac13\alpha_1+\atil_3, \nn
a^{(5)}_{3 }&= \left(\tfrac18-\tfrac38\zeta(3)\right)\lambda, \quad& a^{(5)}_{11 }&= -\tfrac{7}{128}\lambda+\tfrac{5}{24}\alpha_1-\tfrac13\atil_3, \nn
a^{(5)}_{4}&=-\tfrac{145}{82944}\lambda+\tfrac{1}{144}\atil_2+\tfrac{1}{144}\atil_3, \quad& a^{(5)}_{12 }&= \tfrac{107}{96}\lambda+\tfrac32\alpha_1+\atil_1, \nn
a^{(5)}_{5 }&= -\tfrac{5}{41472}\lambda-\tfrac{11}{864}\alpha_1+\tfrac{1}{24}\atil_2+\tfrac{1}{36}\atil_3, \quad& a^{(5)}_{13 }&= -\tfrac{5}{32}\lambda-\tfrac56\alpha_1-\tfrac13\atil_1, \nn
a^{(5)}_{6 }&= \tfrac{29}{2304}\lambda-\tfrac{11}{432}\alpha_1+\tfrac{1}{36}\atil_3, \quad& a^{(5)}_{14 }&= -\tfrac{35}{128}\lambda-\tfrac18\alpha_1-\tfrac16\atil_1+\atil_2, \nn
a^{(5)}_{7}&=\tfrac{1}{72}\atil_2, \quad& a^{(5)}_{15 }&= \tfrac{101}{3456}\lambda+\tfrac{7}{72}\alpha_1+\tfrac{1}{24}\atil_1-\tfrac13\atil_2, \nn
a^{(5)}_{8 }&= -\tfrac{1}{8}\lambda, \quad& a^{(5)}_{16 }&= -\tfrac{5}{2304}\lambda+\tfrac{7}{144}\alpha_1+\tfrac{1}{72}\atil_1,
\end{align}

\begin{table}[h]
	\setlength{\extrarowheight}{0.5cm}
	\setlength{\tabcolsep}{24pt}
	\hspace*{-9.5cm}
	\centering
	\resizebox{6cm}{!}{
		\begin{tabular*}{20cm}{cccccc}
			\begin{picture}(162,203) (287,-190)
			\SetWidth{1.0}
			\SetColor{Black}
			\Line(306,-135)(357,-93)
			\Line(431,-134)(379,-93)
			\Arc(368,-85)(80,127,487)
			\Line(304,-37)(357,-79)
			\Line(430,-35)(379,-79)
			\Line(368,-6)(368,-164)
			\SetWidth{2.0}
			\Line(352,11)(384,-21)
			\Line(384,11)(352,-21)
			\SetWidth{1.0}
			\CTri(345.373,-162)(368,-139.373)(390.627,-162){Black}{Black}\CTri(345.373,-162)(368,-184.627)(390.627,-162){Black}{Black}
			\Text(342,-184)[]{\Huge{\Black{$(1)$}}}
			\end{picture}
			&
			\begin{picture}(186,179) (263,-190)
			\SetWidth{1.0}
			\SetColor{Black}
			\Line(306,-159)(357,-117)
			\Line(431,-158)(379,-117)
			\Arc(368,-109)(80,127,487)
			\Line(304,-61)(357,-103)
			\Line(430,-59)(379,-103)
			\Line(368,-30)(368,-188)
			\SetWidth{2.0}
			\Line(352,-13)(384,-45)
			\Line(384,-13)(352,-45)
			\SetWidth{1.0}
			\CTri(281.373,-158)(304,-135.373)(326.627,-158){Black}{Black}\CTri(281.373,-158)(304,-180.627)(326.627,-158){Black}{Black}
			\Text(278,-180)[]{\Huge{\Black{$(1)$}}}
			\end{picture}
			&
			\begin{picture}(189,179) (260,-190)
			\SetWidth{1.0}
			\SetColor{Black}
			\Line(306,-159)(357,-117)
			\Line(431,-158)(379,-117)
			\Arc(368,-109)(80,127,487)
			\Line(304,-61)(357,-103)
			\Line(430,-59)(379,-103)
			\Line(368,-30)(368,-188)
			\SetWidth{2.0}
			\Line(352,-13)(384,-45)
			\Line(384,-13)(352,-45)
			\SetWidth{1.0}
			\CTri(282.373,-61)(305,-38.373)(327.627,-61){Black}{Black}\CTri(282.373,-61)(305,-83.627)(327.627,-61){Black}{Black}
			\Text(275,-44)[]{\Huge{\Black{$(1)$}}}
			\end{picture}
			&
			\begin{picture}(162,179) (287,-190)
			\SetWidth{1.0}
			\SetColor{Black}
			\Arc(368,-109)(80,127,487)
			\Arc(368,-110)(31.064,123,483)
			\Line(368,-29)(368,-79)
			\Line(345,-131)(304,-157)
			\Line(391,-131)(431,-157)
			\SetWidth{2.0}
			\Line(352,-13)(384,-45)
			\Line(384,-13)(352,-45)
			\SetWidth{1.0}
			\CTri(345.373,-77)(368,-54.373)(390.627,-77){Black}{Black}\CTri(345.373,-77)(368,-99.627)(390.627,-77){Black}{Black}
			\Text(330,-77)[]{\Huge{\Black{$(1)$}}}
			\end{picture}
			&
			\begin{picture}(162,179) (287,-190)
			\SetWidth{1.0}
			\SetColor{Black}
			\Arc(368,-109)(80,127,487)
			\Arc(368,-110)(31.064,123,483)
			\Line(368,-29)(368,-79)
			\Line(345,-131)(304,-157)
			\Line(391,-131)(431,-157)
			\SetWidth{2.0}
			\Line(352,-13)(384,-45)
			\Line(384,-13)(352,-45)
			\SetWidth{1.0}
			\CTri(319.373,-132)(342,-109.373)(364.627,-132){Black}{Black}\CTri(319.373,-132)(342,-154.627)(364.627,-132){Black}{Black}
			\Text(317,-111)[]{\Huge{\Black{$(1)$}}}
			\end{picture}
			&
			\begin{picture}(186,179) (263,-190)
			\SetWidth{1.0}
			\SetColor{Black}
			\Arc(368,-109)(80,127,487)
			\Arc(368,-110)(31.064,123,483)
			\Line(368,-29)(368,-79)
			\Line(345,-131)(304,-157)
			\Line(391,-131)(431,-157)
			\SetWidth{2.0}
			\Line(352,-13)(384,-45)
			\Line(384,-13)(352,-45)
			\SetWidth{1.0}
			\CTri(282.373,-159)(305,-136.373)(327.627,-159){Black}{Black}\CTri(282.373,-159)(305,-181.627)(327.627,-159){Black}{Black}
			\Text(278,-179)[]{\Huge{\Black{$(1)$}}}
			\end{picture}
			\\
			{\Huge $T^{(4)}_{1}$}
			&
			{\Huge $T^{(4)}_{2}$}
			&
			{\Huge $T^{(4)}_{3}$}
			&	
			{\Huge $T^{(4)}_{4}$}
			&
			{\Huge $T^{(4)}_{5}$}
			&
			{\Huge $T^{(4)}_{6}$}
			\\
			&
			&
			&
			&
			&
			\\
			\begin{picture}(168,176) (281,-206)
			\SetWidth{1.0}
			\SetColor{Black}
			\Arc(368,-112)(80,127,487)
			\Arc[clock](269.598,-112.405)(82.403,51.407,-50.512)
			\Arc(466.565,-112)(81.565,128.311,231.689)
			\Line(385,-112)(448,-112)
			\SetWidth{2.0}
			\Line(305,-32)(337,-64)
			\Line(337,-32)(305,-64)
			\SetWidth{1.0}
			\CTri(299.373,-177)(322,-154.373)(344.627,-177){Black}{Black}\CTri(299.373,-177)(322,-199.627)(344.627,-177){Black}{Black}
			\Text(296,-200)[]{\Huge{\Black{$(1)$}}}
			\end{picture}
			&
			\begin{picture}(175,170) (287,-206)
			\SetWidth{1.0}
			\SetColor{Black}
			\Arc(368,-118)(80,127,487)
			\Arc[clock](269.598,-118.405)(82.403,51.407,-50.512)
			\Arc(466.565,-118)(81.565,128.311,231.689)
			\Line(385,-118)(448,-118)
			\SetWidth{2.0}
			\Line(401,-38)(433,-70)
			\Line(433,-38)(401,-70)
			\SetWidth{1.0}
			\CTri(392.373,-182)(415,-159.373)(437.627,-182){Black}{Black}\CTri(392.373,-182)(415,-204.627)(437.627,-182){Black}{Black}
			\Text(445,-199)[]{\Huge{\Black{$(1)$}}}
			\end{picture}
			&
			\begin{picture}(179,162) (287,-207)
			\SetWidth{1.0}
			\SetColor{Black}
			\Arc(368,-126)(80,127,487)
			\Arc[clock](269.598,-126.405)(82.403,51.407,-50.512)
			\Arc(466.565,-126)(81.565,128.311,231.689)
			\Line(385,-126)(448,-126)
			\SetWidth{2.0}
			\Line(432,-109)(464,-141)
			\Line(464,-109)(432,-141)
			\SetWidth{1.0}
			\CTri(365.373,-126)(388,-103.373)(410.627,-126){Black}{Black}\CTri(365.373,-126)(388,-148.627)(410.627,-126){Black}{Black}
			\Text(369,-99)[]{\Huge{\Black{$(1)$}}}
			\end{picture}
			&
			\begin{picture}(162,202) (287,-190)
			\SetWidth{1.0}
			\SetColor{Black}
			\Arc(368,-86)(80,127,487)
			\Arc[clock](269.598,-86.405)(82.403,51.407,-50.512)
			\Arc(466.565,-86)(81.565,128.311,231.689)
			\Line(368,-6)(368,-166)
			\SetWidth{2.0}
			\Line(352,10)(384,-22)
			\Line(384,10)(352,-22)
			\SetWidth{1.0}
			\CTri(345.373,-163)(368,-140.373)(390.627,-163){Black}{Black}\CTri(345.373,-163)(368,-185.627)(390.627,-163){Black}{Black}
			\Text(338,-184)[]{\Huge{\Black{$(1)$}}}
			\end{picture}
			&
			\begin{picture}(170,175) (279,-206)
			\SetWidth{1.0}
			\SetColor{Black}
			\Arc(368,-113)(80,127,487)
			\Arc[clock](269.598,-113.405)(82.403,51.407,-50.512)
			\Arc(466.565,-113)(81.565,128.311,231.689)
			\Line(368,-33)(368,-193)
			\SetWidth{2.0}
			\Line(305,-33)(337,-65)
			\Line(337,-33)(305,-65)
			\SetWidth{1.0}
			\CTri(297.373,-177)(320,-154.373)(342.627,-177){Black}{Black}\CTri(297.373,-177)(320,-199.627)(342.627,-177){Black}{Black}
			\Text(294,-200)[]{\Huge{\Black{$(1)$}}}
			\end{picture}
			&
			\begin{picture}(173,170) (287,-199)
			\SetWidth{1.0}
			\SetColor{Black}
			\Arc(368,-118)(80,127,487)
			\Arc[clock](269.598,-118.405)(82.403,51.407,-50.512)
			\Arc(466.565,-118)(81.565,128.311,231.689)
			\Line(368,-38)(368,-198)
			\SetWidth{2.0}
			\Line(305,-38)(337,-70)
			\Line(337,-38)(305,-70)
			\SetWidth{1.0}
			\CTri(393.373,-55)(416,-32.373)(438.627,-55){Black}{Black}\CTri(393.373,-55)(416,-77.627)(438.627,-55){Black}{Black}
			\Text(443,-39)[]{\Huge{\Black{$(1)$}}}
			\end{picture}
			\\
			{\Huge $T^{(4)}_{7}$}
			&
			{\Huge $T^{(4)}_{8}$}
			&
			{\Huge $T^{(4)}_{9}$}
			&	
			{\Huge $T^{(4)}_{10}$}
			&
			{\Huge $T^{(4)}_{11}$}
			&
			{\Huge $T^{(4)}_{12}$}
			\\
			&
			&
			&
			&
			&
			\\
			\begin{picture}(173,175) (287,-206)
			\SetWidth{1.0}
			\SetColor{Black}
			\Arc(368,-113)(80,127,487)
			\Arc[clock](269.598,-113.405)(82.403,51.407,-50.512)
			\Arc(466.565,-113)(81.565,128.311,231.689)
			\Line(368,-33)(368,-193)
			\SetWidth{2.0}
			\Line(305,-33)(337,-65)
			\Line(337,-33)(305,-65)
			\SetWidth{1.0}
			\CTri(393.373,-177)(416,-154.373)(438.627,-177){Black}{Black}\CTri(393.373,-177)(416,-199.627)(438.627,-177){Black}{Black}
			\Text(443,-200)[]{\Huge{\Black{$(1)$}}}
			\end{picture}
			&
			\begin{picture}(197,171) (252,-198)
			\SetWidth{1.0}
			\SetColor{Black}
			\Arc(368,-117)(80,127,487)
			\Arc[clock](367.5,-181.522)(43.525,-178.054,-361.946)
			\Arc(317.159,-84.339)(40.18,-134.494,68.324)
			\Arc[clock](421.391,-84.57)(39.676,-47.882,-247.176)
			\SetWidth{2.0}
			\Line(315,-29)(347,-61)
			\Line(347,-29)(315,-61)
			\SetWidth{1.0}
			\CTri(268.373,-118)(291,-95.373)(313.627,-118){Black}{Black}\CTri(268.373,-118)(291,-140.627)(313.627,-118){Black}{Black}
			\Text(267,-98)[]{\Huge{\Black{$(1)$}}}
			\end{picture}
			&
			\begin{picture}(162,181) (287,-198)
			\SetWidth{1.0}
			\SetColor{Black}
			\Arc(368,-107)(80,127,487)
			\Arc[clock](367.5,-171.522)(43.525,-178.054,-361.946)
			\Arc(317.159,-74.339)(40.18,-134.494,68.324)
			\Arc[clock](421.391,-74.57)(39.676,-47.882,-247.176)
			\SetWidth{2.0}
			\Line(315,-19)(347,-51)
			\Line(347,-19)(315,-51)
			\SetWidth{1.0}
			\CTri(381.373,-172)(404,-149.373)(426.627,-172){Black}{Black}\CTri(381.373,-172)(404,-194.627)(426.627,-172){Black}{Black}
			\Text(431,-192)[]{\Huge{\Black{$(1)$}}}
			\end{picture}
			&
			\begin{picture}(167,171) (287,-198)
			\SetWidth{1.0}
			\SetColor{Black}
			\Arc(368,-117)(80,127,487)
			\Arc[clock](367.5,-181.522)(43.525,-178.054,-361.946)
			\Arc(317.159,-84.339)(40.18,-134.494,68.324)
			\Arc[clock](421.391,-84.57)(39.676,-47.882,-247.176)
			\SetWidth{2.0}
			\Line(315,-29)(347,-61)
			\Line(347,-29)(315,-61)
			\SetWidth{1.0}
			\CTri(385.373,-51)(408,-28.373)(430.627,-51){Black}{Black}\CTri(385.373,-51)(408,-73.627)(430.627,-51){Black}{Black}
			\Text(437,-40)[]{\Huge{\Black{$(1)$}}}
			\end{picture}
			&
			\begin{picture}(197,171) (287,-198)
			\SetWidth{1.0}
			\SetColor{Black}
			\Arc(368,-117)(80,127,487)
			\Arc[clock](367.5,-181.522)(43.525,-178.054,-361.946)
			\Arc(317.159,-84.339)(40.18,-134.494,68.324)
			\Arc[clock](421.391,-84.57)(39.676,-47.882,-247.176)
			\SetWidth{2.0}
			\Line(315,-29)(347,-61)
			\Line(347,-29)(315,-61)
			\SetWidth{1.0}
			\CTri(422.373,-116)(445,-93.373)(467.627,-116){Black}{Black}\CTri(422.373,-116)(445,-138.627)(467.627,-116){Black}{Black}
			\Text(467,-140)[]{\Huge{\Black{$(1)$}}}
			\end{picture}
			&
			\begin{picture}(166,182) (283,-198)
			\SetWidth{1.0}
			\SetColor{Black}
			\Arc(368,-106)(80,127,487)
			\Arc[clock](367.5,-170.522)(43.525,-178.054,-361.946)
			\Arc(317.159,-73.339)(40.18,-134.494,68.324)
			\Arc[clock](421.391,-73.57)(39.676,-47.882,-247.176)
			\SetWidth{2.0}
			\Line(315,-18)(347,-50)
			\Line(347,-18)(315,-50)
			\SetWidth{1.0}
			\CTri(306.373,-172)(329,-149.373)(351.627,-172){Black}{Black}\CTri(306.373,-172)(329,-194.627)(351.627,-172){Black}{Black}
			\Text(298,-192)[]{\Huge{\Black{$(1)$}}}
			\end{picture}
			\\
			{\Huge $T^{(4)}_{13}$}
			&
			{\Huge $T^{(4)}_{14}$}
			&
			{\Huge $T^{(4)}_{15}$}
			&	
			{\Huge $T^{(4)}_{16}$}
			&
			{\Huge $T^{(4)}_{17}$}
			&
			{\Huge $T^{(4)}_{18}$}
			\\
			&
			&
			&
			&
			&
			\\
			\begin{picture}(171,167) (287,-202)
			\SetWidth{1.0}
			\SetColor{Black}
			\Arc(368,-121)(80,127,487)
			\Arc[clock](269.598,-121.405)(82.403,51.407,-50.512)
			\Arc(466.565,-121)(81.565,128.311,231.689)
			\Line(385,-121)(448,-121)
			\SetWidth{2.0}
			\Line(305,-41)(337,-73)
			\Line(337,-41)(305,-73)
			\SetWidth{1.0}
			\CTri(393.373,-59)(416,-36.373)(438.627,-59){Black}{Black}\CTri(393.373,-59)(416,-81.627)(438.627,-59){Black}{Black}
			\Text(441,-45)[]{\Huge{\Black{$(1)$}}}
			\end{picture}
			&
			\begin{picture}(173,168) (276,-201)
			\SetWidth{1.0}
			\SetColor{Black}
			\Arc(368,-120)(80,127,487)
			\Arc[clock](269.598,-120.405)(82.403,51.407,-50.512)
			\Arc(466.565,-120)(81.565,128.311,231.689)
			\Line(385,-120)(448,-120)
			\SetWidth{2.0}
			\Line(401,-40)(433,-72)
			\Line(433,-40)(401,-72)
			\SetWidth{1.0}
			\CTri(298.373,-58)(321,-35.373)(343.627,-58){Black}{Black}\CTri(298.373,-58)(321,-80.627)(343.627,-58){Black}{Black}
			\Text(291,-43)[]{\Huge{\Black{$(1)$}}}
			\end{picture}
			&
			\begin{picture}(195,163) (287,-206)
			\SetWidth{1.0}
			\SetColor{Black}
			\Arc(368,-125)(80,127,487)
			\Arc[clock](269.598,-125.405)(82.403,51.407,-50.512)
			\Arc(466.565,-125)(81.565,128.311,231.689)
			\Line(385,-125)(448,-125)
			\SetWidth{2.0}
			\Line(305,-45)(337,-77)
			\Line(337,-45)(305,-77)
			\SetWidth{1.0}
			\CTri(422.373,-125)(445,-102.373)(467.627,-125){Black}{Black}\CTri(422.373,-125)(445,-147.627)(467.627,-125){Black}{Black}
			\Text(465,-152)[]{\Huge{\Black{$(1)$}}}
			\end{picture}
			&
			\begin{picture}(191,168) (275,-201)
			\SetWidth{1.0}
			\SetColor{Black}
			\Arc(368,-120)(80,127,487)
			\Arc[clock](269.598,-120.405)(82.403,51.407,-50.512)
			\Arc(466.565,-120)(81.565,128.311,231.689)
			\Line(385,-120)(448,-120)
			\SetWidth{2.0}
			\Line(432,-103)(464,-135)
			\Line(464,-103)(432,-135)
			\SetWidth{1.0}
			\CTri(297.373,-57)(320,-34.373)(342.627,-57){Black}{Black}\CTri(297.373,-57)(320,-79.627)(342.627,-57){Black}{Black}
			\Text(290,-43)[]{\Huge{\Black{$(1)$}}}
			\end{picture}
			&
			\begin{picture}(167,178) (287,-206)
			\SetWidth{1.0}
			\SetColor{Black}
			\Arc(368,-110)(80,127,487)
			\Arc[clock](269.598,-110.405)(82.403,51.407,-50.512)
			\Arc(466.565,-110)(81.565,128.311,231.689)
			\Line(385,-110)(448,-110)
			\SetWidth{2.0}
			\Line(305,-30)(337,-62)
			\Line(337,-30)(305,-62)
			\SetWidth{1.0}
			\CTri(394.373,-173)(417,-150.373)(439.627,-173){Black}{Black}\CTri(394.373,-173)(417,-195.627)(439.627,-173){Black}{Black}
			\Text(437,-200)[]{\Huge{\Black{$(1)$}}}
			\end{picture}
			&
			\begin{picture}(173,172) (276,-206)
			\SetWidth{1.0}
			\SetColor{Black}
			\Arc(368,-116)(80,127,487)
			\Arc[clock](269.598,-116.405)(82.403,51.407,-50.512)
			\Arc(466.565,-116)(81.565,128.311,231.689)
			\Line(385,-116)(448,-116)
			\SetWidth{2.0}
			\Line(401,-36)(433,-68)
			\Line(433,-36)(401,-68)
			\SetWidth{1.0}
			\CTri(298.373,-182)(321,-159.373)(343.627,-182){Black}{Black}\CTri(298.373,-182)(321,-204.627)(343.627,-182){Black}{Black}
			\Text(291,-200)[]{\Huge{\Black{$(1)$}}}
			\end{picture}
			\\
			{\Huge $T^{(4)}_{19}$}
			&
			{\Huge $T^{(4)}_{20}$}
			&
			{\Huge $T^{(4)}_{21}$}
			&	
			{\Huge $T^{(4)}_{22}$}
			&
			{\Huge $T^{(4)}_{23}$}
			&
			{\Huge $T^{(4)}_{24}$}
			\\
			&
			&
			&
			&
			&
			\\
			\begin{picture}(194,163) (287,-206)
			\SetWidth{1.0}
			\SetColor{Black}
			\Arc(368,-125)(80,127,487)
			\Arc[clock](269.598,-125.405)(82.403,51.407,-50.512)
			\Arc(466.565,-125)(81.565,128.311,231.689)
			\Line(385,-125)(448,-125)
			\SetWidth{2.0}
			\Line(401,-45)(433,-77)
			\Line(433,-45)(401,-77)
			\SetWidth{1.0}
			\CTri(422.373,-125)(445,-102.373)(467.627,-125){Black}{Black}\CTri(422.373,-125)(445,-147.627)(467.627,-125){Black}{Black}
			\Text(464,-152)[]{\Huge{\Black{$(1)$}}}
			\end{picture}
			&
			\begin{picture}(179,169) (287,-200)
			\SetWidth{1.0}
			\SetColor{Black}
			\Arc(368,-119)(80,127,487)
			\Arc[clock](269.598,-119.405)(82.403,51.407,-50.512)
			\Arc(466.565,-119)(81.565,128.311,231.689)
			\Line(385,-119)(448,-119)
			\SetWidth{2.0}
			\Line(432,-102)(464,-134)
			\Line(464,-102)(432,-134)
			\SetWidth{1.0}
			\CTri(393.373,-56)(416,-33.373)(438.627,-56){Black}{Black}\CTri(393.373,-56)(416,-78.627)(438.627,-56){Black}{Black}
			\Text(445,-41)[]{\Huge{\Black{$(1)$}}}
			\end{picture}
			&
			\begin{picture}(173,179) (276,-190)
			\SetWidth{1.0}
			\SetColor{Black}
			\Arc(368,-109)(80,127,487)
			\Arc[clock](269.598,-109.405)(82.403,51.407,-50.512)
			\Arc(466.565,-109)(81.565,128.311,231.689)
			\Line(368,-29)(368,-189)
			\SetWidth{2.0}
			\Line(352,-13)(384,-45)
			\Line(384,-13)(352,-45)
			\SetWidth{1.0}
			\CTri(297.373,-45)(320,-22.373)(342.627,-45){Black}{Black}\CTri(297.373,-45)(320,-67.627)(342.627,-45){Black}{Black}
			\Text(291,-29)[]{\Huge{\Black{$(1)$}}}
			\end{picture}
			&
			\begin{picture}(162,184) (287,-185)
			\SetWidth{1.0}
			\SetColor{Black}
			\Arc(368,-104)(80,127,487)
			\Arc[clock](269.598,-104.405)(82.403,51.407,-50.512)
			\Arc(466.565,-104)(81.565,128.311,231.689)
			\Line(368,-24)(368,-184)
			\SetWidth{2.0}
			\Line(305,-24)(337,-56)
			\Line(337,-24)(305,-56)
			\SetWidth{1.0}
			\CTri(345.373,-27)(368,-4.373)(390.627,-27){Black}{Black}\CTri(345.373,-27)(368,-49.627)(390.627,-27){Black}{Black}
			\Text(395,-11)[]{\Huge{\Black{$(1)$}}}
			\end{picture}
			&
			\begin{picture}(171,189) (278,-190)
			\SetWidth{1.0}
			\SetColor{Black}
			\Arc(368,-99)(80,127,487)
			\Arc[clock](269.598,-99.405)(82.403,51.407,-50.512)
			\Arc(466.565,-99)(81.565,128.311,231.689)
			\Line(368,-19)(368,-179)
			\SetWidth{2.0}
			\Line(352,-3)(384,-35)
			\Line(384,-3)(352,-35)
			\SetWidth{1.0}
			\CTri(297.373,-163)(320,-140.373)(342.627,-163){Black}{Black}\CTri(297.373,-163)(320,-185.627)(342.627,-163){Black}{Black}
			\Text(293,-184)[]{\Huge{\Black{$(1)$}}}
			\end{picture}
			&
			\begin{picture}(162,188) (287,-206)
			\SetWidth{1.0}
			\SetColor{Black}
			\Arc(368,-100)(80,127,487)
			\Arc[clock](269.598,-100.405)(82.403,51.407,-50.512)
			\Arc(466.565,-100)(81.565,128.311,231.689)
			\Line(368,-20)(368,-180)
			\SetWidth{2.0}
			\Line(305,-20)(337,-52)
			\Line(337,-20)(305,-52)
			\SetWidth{1.0}
			\CTri(345.373,-177)(368,-154.373)(390.627,-177){Black}{Black}\CTri(345.373,-177)(368,-199.627)(390.627,-177){Black}{Black}
			\Text(343,-200)[]{\Huge{\Black{$(1)$}}}
			\end{picture}
			\\
			{\Huge $T^{(4)}_{25}$}
			&
			{\Huge $T^{(4)}_{26}$}
			&
			{\Huge $T^{(4)}_{27}$}
			&	
			{\Huge $T^{(4)}_{28}$}
			&
			{\Huge $T^{(4)}_{29}$}
			&
			{\Huge $T^{(4)}_{30}$}
		\end{tabular*}
	}
	\caption{Next-to-next-to-leading-order metric terms $T^{(4)}$}
	\label{Tdef}	
\end{table}
where $\lambda$ is again defined in Eq.~\eqref{Athreea} and we explicitly display the freedom as expressed in Eq.~\eqref{free}
\begin{align}
A\rightarrow& A+2\alpha_1\beta^{(1)ijk}\beta^{(2)ijk}+\atil_1\beta^{(1)ijk}\beta^{(1)klm}g_{22}^{iljm}
+\atil_2\beta^{(1)ikl}\beta^{(1)jkl}g^{ij}_{(1A)}\nn
&+\atil_3\beta^{(1)ijk}\beta^{(1)lmn}g^{jkl}g^{imn},
\label{afree}
\end{align}
using the results of Eqs.~\eqref{betatwo}, \eqref{cmstwo}. The terms with $\atil_1$-$\atil_3$ correspond to the terms with
$T^{(3)}_1$-$T^{(3)}_3$ in Table~\ref{Tmetric} (note that $T^{(3)}_3$, $T^{(3)}_4$ have the same effect at this order, as may be observed in Eq.~\eqref{tcoeffs}).

At this loop order the tensor $T_{IJ}$  is {\it not} inevitably symmetric. Analogously to  the previous order, the metric may be expressed in the form
\be
T_{ijk,lmn}^{(4)}=\sum_{\alpha=1}^{30}t^{(4)}_{\alpha}\left(T_{\alpha}^{(4)}\right)_{ijk,lmn}.
\label{metfour}
\ee
The structures which may arise are depicted in
Table~\ref{Tdef}, using a similar convention to the previous order. As before, a cross denotes $(dg)^{ijk}$ and a diamond represents $\beta^{ijk}$. We see immediately that $T^{(4)}_{1}\ldots T^{(4)}_{16}$ are individually symmetric; the remaining diagrams are grouped in pairs whose coefficients should be equal for symmetry.
Solving Eq.~\eqref{grad} certainly does not guarantee that $t^{(4)}_{17}=t^{(4)}_{18}$, $t^{(4)}_{19}=t^{(4)}_{20}$
etc. However it turns out that we can impose symmetry on $T^{(4)}_{IJ}$ provided the additional condition
\be
 c_{(3m)}-2c_{(3n)}+\tfrac16c_{(3E)}-c_{(3H)}-12c_{(3L)}-Z
=\tfrac{11}{36}(c_{(2B)}-2c_{(2c)})
\label{symcons}
\ee
 is satisfied. The values in Eq.~\eqref{betthree}, \eqref{betthreea} do indeed satisfy this condition and we obtain a symmetric metric with the values

\begin{align}
t^{(4)}_{1} &= \tfrac{13}{8}\lambda-\tfrac{3}{2}\lambda \zeta(3)-3\alpha_1-t^{(4)}_{3},\nn
t^{(4)}_{2} &= \tfrac{1}{4}\lambda-2\alpha_1,\nn
t^{(4)}_{4} &= -\tfrac{161}{48}\lambda+\tfrac{11}{2}\alpha_1+24\atil_2,\nn
t^{(4)}_{5} &= -\tfrac{89}{24}\lambda+11\alpha_1+48\atil_2,\nn
t^{(4)}_{6} &= -\tfrac{31}{24}\lambda +4\alpha_1+24\atil_2,\nn
t^{(4)}_{7} &= -\tfrac{13}{24}\lambda+\tfrac13\alpha_1-2\atil_2,\nn
t^{(4)}_{8} &= \tfrac{49}{144}\lambda-\tfrac{7}{12}\alpha_1-4\atil_2,\nn
t^{(4)}_{9} &= -\tfrac{11}{96}\lambda-2\atil_2,\nn
t^{(4)}_{10} &= \tfrac13\atil_2,\nn
t^{(4)}_{11} &= \tfrac{391}{1728}\lambda-\tfrac{11}{72}\alpha_1+\tfrac13\atil_2,\nn
t^{(4)}_{12} &= \tfrac{11}{432}\lambda+\tfrac43\atil_2-t^{(4)}_{13},\nn
t^{(4)}_{14} &= \tfrac{1}{192}\lambda+\tfrac13\atil_2,\nn
t^{(4)}_{15} &= -\tfrac{299}{1728}\lambda-t^{(4)}_{16}-2t^{(4)}_{17}+\tfrac{11}{36}\alpha_1+\tfrac53\atil_2,\nn
t^{(4)}_{17}&=t^{(4)}_{18},\nn
t^{(4)}_{19} &=t^{(4)}_{20}= -\tfrac{59}{72}\lambda+\tfrac23t^{(3)}_{4}-12(t^{(4)}_{16}+t^{(4)}_{17})+\tfrac{11}{12}\alpha_1+6\atil_2,\nn
t^{(4)}_{21} &=t^{(4)}_{22}= \tfrac{115}{288}\lambda-\tfrac76\alpha_1-8\atil_2,\nn
t^{(4)}_{23} &=t^{(4)}_{24}=\tfrac{73}{48}\lambda-\tfrac23t^{(3)}_{4}+12(t^{(4)}_{16}+t^{(4)}_{17})-\tfrac{25}{12}\alpha_1-12\atil_2,\nn
t^{(4)}_{25} &=t^{(4)}_{26}= \tfrac{101}{288}\lambda-\tfrac{7}{12}\alpha_1-6\atil_2,\nn
t^{(4)}_{27} &=t^{(4)}_{28} =\tfrac{373}{1728}\lambda-\tfrac{11}{36}t^{(3)}_{4}+2(t^{(4)}_{16}+t^{(4)}_{17})-\tfrac{11}{72}\alpha_1-\atil_2,\nn
t^{(4)}_{29} &=t^{(4)}_{30}=-\tfrac{11}{48}\lambda+\tfrac{11}{36}t^{(3)}_{4}-2(t^{(4)}_{16}+t^{(4)}_{17})+\tfrac{11}{72}\alpha_1+2\atil_2.
\label{symvals}
\end{align}
The values of $t^{(4)}_3$, $t^{(4)}_{13}$, $t^{(4)}_{16}$, $t^{(4)}_{17}$
remain arbitrary, in a similar fashion to the previous order where in Eq.~\eqref{tcoeffs} only $t^{(3)}_3+t^{(3)}_4$ was determined. Before imposing symmetry, the $T^{(4)}_{IJ}$ coefficients would also display the freedom Eq.~\eqref{free} as expressed at this order in Eq.~\eqref{afree}; however, the general redefinition Eq.~\eqref{afree} is not compatible with symmetry of $T^{(4)}_{IJ}$. Nevertheless, as may be seen in Eq.~\eqref{symvals}, there is still a residual two-parameter freedom expressed in $\alpha_1$, $\atil_2$, corresponding to choosing
\be
\atil_1=-12\atil_2-\tfrac{7}{2}\alpha_1,\quad \atil_3=2\atil_2+\tfrac{11}{24}\alpha_1.
\ee
It is worth remarking that the freedom in $\alpha_1$ is in general only preserved in the symmetric case providing Eq.~\eqref{constwo} is satisfied and is therefore somewhat non-trivial. We note that in the four-dimensional case, the requirement of symmetry of $T_{IJ}$ was more restrictive and was not possible within $\MSbar$\cite{OsbJacnew}.

\section{Scheme changes}

In this section we shall turn to a fuller discussion of scheme changes such as that from $\MSbar$ to MOM.
As we mentioned in Sect. 3, we have obtained the three-loop MOM $\beta$-function by implementing the appropriate scheme change, avoiding a separate three-loop Feynman diagram calculation for MOM. Here we wish to consider the effect of more general scheme changes, in order to demonstrate the scheme-invariance of  the consistency conditions on the $\beta$-function coefficients, Eqs.~\eqref{conthree}; we shall therefore give our results in full generality. We now rewrite the coupling redefinition of Eq.~\eqref{mapexp}, which implements the change of scheme, in the form
\be g^{ijk}\rightarrow \gbar^{ijk}\equiv \gbar^{ ijk}(g),
\label{genredef}
\ee
returning to the general couplings of Eq.~\eqref{lag} and a general scheme change.
The effects of Eq.~\eqref{genredef} may be computed from the generalisation of Eq.~\eqref{betarel},
\be
\overline{\beta}^{ijk}(\gbar)=\mu\frac{d}{d\mu}\gbar^{ ijk}=\beta \cdot \frac{\pa}{\pa g}\gbar^{ ijk}(g)
\label{redef}
\ee
 (where $\beta \cdot \frac{\pa}{\pa g}\equiv\beta^{klm}\frac{\pa}{\pa g^{klm}}$) which to lowest order may be written
  \be
  \delta \beta^{ijk}=\beta \cdot \frac{\pa}{\pa g}\delta g^{ijk} -\delta g \cdot \frac{\pa}{\pa g}\beta^{ijk}.
\label{redefa}
\ee
The effect of a one-loop change
\be
\delta g=\delta_1 g_{(1a)}+\delta_2 g_{(1A)}
\label{redonea}
\ee
on the two-loop $\beta$-functions is easily computed as
\be \delta c_{(2B)}=-\tfrac16\Delta, \quad \delta c_{(2c)}=\tfrac16\Delta,\quad
\Delta=\delta_1+12\delta_2
\label{redone}
\ee
where $c_{(2B)}$ and $c_{(2c)}$ are defined in Eq.~\eqref{betatwo}. It is readily checked that the consistency 
condition  Eq.~\eqref{constwo} is invariant under Eq.~\eqref{redone}, as expected.

At three loops we consider redefinitions
\be
\delta g =\epsilon_1 g_{(2b)} +\epsilon_2 g_{(2c)}+ \epsilon_3g_{(2d)}+\epsilon_4g_{(2e)}+\epsilon_5g_{(2f)}+\epsilon_6 g_{(2B)}
+\epsilon_7 g_{(2C)}+\epsilon_8g_{(2D)},
\label{redefth}
\ee

\noindent where

 \be
g_{(2e)}=g^{ilm}g_{(1A)}^{lj}g_{(1A)}^{mk},
 \quad g_{(2f)}^{ijk}=g_{(1a)}^{ijl}g_{(1A)}^{lk},
 \quad g_{(2D)}^{ijk}=g^{ijl}g_{(1A)}^{lm}g_{(1A)}^{mk}
\label{tensb}
\ee

\begin{table}[h]
	\setlength{\extrarowheight}{1cm}
	\setlength{\tabcolsep}{24pt}
	\hspace*{-1.5cm}
	\centering
	\resizebox{7.5cm}{!}{
		\begin{tabular*}{20cm}{ccccc}
			\resizebox{6cm}{!}{\begin{picture}(135,118) (267,-235)
				\SetWidth{1.0}
				\SetColor{Black}
				\Line(315,-186)(303,-168)
				\Line(315,-187)(287,-234)
				\Line(379,-186)(401,-186)
				\Arc(359,-186)(20,233,593)
				\Line(316,-186)(338,-186)
				\Arc(292,-150)(20,233,593)
				\Line(268,-118)(279,-134)
				\end{picture}}
			&
			\resizebox{6cm}{!}{\begin{picture}(139,111) (299,-247)
				\SetWidth{1.0}
				\SetColor{Black}
				\Arc(336,-193)(35.777,153,513)
				\Line(371,-193)(390,-193)
				\Line(316,-163)(300,-137)
				\Line(316,-223)(303,-246)
				\Arc(406,-193)(15.62,140,500)
				\Line(422,-193)(437,-193)
				\end{picture}}
			&
			\resizebox{6cm}{!}{\resizebox{6cm}{!}{\begin{picture}(152,70) (298,-295)
					\SetWidth{1.0}
					\SetColor{Black}
					\Line(379,-252)(391,-252)
					\Line(315,-252)(299,-226)
					\Line(315,-253)(302,-276)
					\Arc(359,-252)(20,233,593)
					\Line(316,-252)(338,-252)
					\Arc(412,-252)(20,233,593)
					\Line(449,-252)(433,-252)
					\end{picture}}}
			\\
			{\Huge $g_{(2e)}$}
			&
			{\Huge $g_{(2f)}$}
			&
			{\Huge $g_{(2D)}$}			
		\end{tabular*}
	}
	\caption{Two-loop 1PR structures arising from coupling redefinitions}
	\label{fig3}	
\end{table}

\noindent as depicted in Table~\ref{fig3},
and the remaining tensor structures are defined in Eqs.~\eqref{tensa};
and we also need to consider the effect at this order of the lower-order redefinitions given by Eqs.~\eqref{redonea}, \eqref{redone}.

In general, in addition to modifying the coefficients already present in the $\MSbar$ $\beta$-function as defined in Eq.~\eqref{betathree} (which correspond to 1PI diagrams, or 1PI wave-function renormalisation diagrams attached as in Eq.~\eqref{anomdim}) these redefinitions will generate tensor structures corresponding to 1PR diagrams given by
\begin{align}
g^{ijk}_{(3\alpha)}=g_{(1A)}^{il}g_{(2b)}^{klj},\quad
g^{ijk}_{(3\beta)}=g_{(1A)}^{jl}g_{(2c)}^{lik},\quad&
g^{ijk}_{(3\gamma)}=g_{(1A)}^{il}g_{(2c)}^{jlk},\quad
g^{ijk}_{(3\delta)}=g_{(2B)}^{il}g_{(1a)}^{ljk},\nn
g^{ijk}_{(3\epsilon)}=g_{(2C)}^{il}g_{(1a)}^{ljk},\quad
g^{ijk}_{(3\zeta)}=g_{(1A)}^{il}g_{(1A)}^{jm}g_{(1a)}^{lmk},\quad&
g^{ijk}_{(3\eta)}=g_{(2D)}^{il}g_{(1a)}^{ljk},\quad
g^{ijk}_{(3\kappa)}=g_{(1A)}^{il}g_{(2B)}^{jm}g^{lmk},\nn
g^{ijk}_{(3\lambda)}=g_{(1A)}^{im}g_{(2C)}^{jl}g^{lmk},\quad
g^{ijk}_{(3\mu)}&=g_{(1A)}^{il}g_{(2D)}^{jm}g^{lmk},\quad
g^{ijk}_{(3\nu)}=g_{(1A)}^{il}g_{(2D)}^{lm}g^{mjk},
\end{align}
the tensor structures on the right-hand sides again defined in Eqs.~\eqref{tensa}. These 1PR structures are depicted in Table.~\ref{onepr}; we denote the coefficients of these tensor structures in the $\beta$-function by  $c_{3\alpha}\ldots c_{3\chi}$, respectively.

We obtain using Eq.~\eqref{redef} (expanding now to 2nd order in $\delta_1$, $\delta_2$ where necessary)
\begin{align}
\delta c_{(3g)}&=-2\epsilon_1-2c_{(2b)}\delta_1+\delta_1^2,\nn
\delta c_{(3h)}&=\tfrac16\epsilon_1-\epsilon_2-c_{(2c)}\delta_1-2c_{(2b)}\delta_2
-\tfrac16\delta_1^2,\nn
\delta c_{(3i)}&=\tfrac13\epsilon_1-2\epsilon_2+2\epsilon_5-2c_{(2c)}\delta_1-4c_{(2b)}\delta_2
-\tfrac13\delta_1^2-2\delta_1\delta_2,\nn
\delta c_{(3j)}&=-2\epsilon_2+2\epsilon_6-2c_{(2c)}\delta_1+2c_{(2B)}\delta_1
-\tfrac13\delta_1^2-4\delta_1\delta_2,\nn
\delta c_{(3k)}&=\tfrac13\epsilon_1+2\epsilon_2+2c_{(2c)}\delta_1-4c_{(2b)}\delta_2,\nn
\delta c_{(3l)}&=\tfrac16\epsilon_1+\epsilon_2+c_{(2c)}\delta_1-2c_{(2b)}\delta_2,\nn
\delta c_{(3m)}&=\tfrac13 \epsilon_2+2\epsilon_7-4c_{(2c)}\delta_2+2c_{(2C)}\delta_1
-2(\tfrac13\delta_1\delta_2+4\delta_2^2),\nn
 \delta c_{(3n)}&=\tfrac13 \epsilon_2+2\epsilon_8-4c_{(2c)}\delta_2
-\tfrac23\delta_1\delta_2-7\delta_2^2,\nn
\delta c_{(3o)}&=\tfrac13 \epsilon_2+\epsilon_4-4c_{(2c)}\delta_2
-\tfrac23\delta_1\delta_2-5\delta_2^2,\nn
\delta c_{(3p)}&=\tfrac13 \epsilon_3-4c_{(2d)}\delta_2,\quad
 \delta c_{(3q)}=-\epsilon_3-c_{(2d)}\delta_1,\nn
\delta c_{(3r)}&=-2\epsilon_3-2c_{(2d)}\delta_1,\quad
\delta c_{(3s)}=\epsilon_3+c_{(2d)}\delta_1,
\label{reddefsa}
\end{align}

\begin{table}[h]
	\setlength{\extrarowheight}{0.5cm}
	\setlength{\tabcolsep}{24pt}
	\hspace*{-6cm}
	\centering
	\resizebox{8cm}{!}{
		\begin{tabular*}{20cm}{cccc}
			\resizebox{7cm}{!}{\begin{picture}(254,218) (224,-191)
				\SetWidth{1.0}
				\SetColor{Black}
				\Arc(261,-169)(19.849,139,499)
				\Line(350,26)(350,-57)
				\Arc(350,-62.155)(36.845,-167.707,-12.293)
				\Arc(350,-118)(60.803,144,504)
				\Line(477,-190)(402,-149)
				\Line(225,-190)(244,-179)
				\Line(280,-159)(297,-149)
				\end{picture}}
			&
			\resizebox{7cm}{!}{\begin{picture}(254,218) (224,-191)
				\SetWidth{1.0}
				\SetColor{Black}
				\Arc[clock](350,-173.845)(36.845,167.707,12.293)
				\Arc(350,-118)(60.803,144,504)
				\Line(225,-190)(297,-149)
				\Line(402,-149)(477,-190)
				\Arc(350,-15)(20.518,133,493)
				\Line(350,26)(350,6)
				\Line(350,-36)(350,-57)
				\end{picture}}
			&
			\resizebox{7cm}{!}{\begin{picture}(254,218) (224,-191)
				\SetWidth{1.0}
				\SetColor{Black}
				\Arc(261,-169)(19.849,139,499)
				\Line(350,26)(350,-57)
				\Arc[clock](350,-173.845)(36.845,167.707,12.293)
				\Arc(350,-118)(60.803,144,504)
				\Line(477,-190)(402,-149)
				\Line(225,-190)(244,-179)
				\Line(280,-159)(297,-149)
				\end{picture}}
			&
			\resizebox{7cm}{!}{\begin{picture}(254,218) (224,-191)
				\SetWidth{1.0}
				\SetColor{Black}
				\Arc(350,-118)(60.803,144,504)
				\Line(225,-190)(297,-149)
				\Line(402,-149)(477,-190)
				\Arc(350,-15)(19.799,135,495)
				\Line(350,26)(350,6)
				\Line(350,-36)(350,-57)
				\Line(330,-15)(370,-15)
				\end{picture}}
			\\
			{\Huge $g^{ijk}_{(3\alpha)}$}
			&
			{\Huge $g^{ijk}_{(3\beta)}$}
			&
			{\Huge $g^{ijk}_{(3\gamma)}$}
			&	
			{\Huge $g^{ijk}_{(3\delta)}$}
			\\
			&
			&
			&
			\\
			\resizebox{7cm}{!}{\begin{picture}(254,218) (224,-191)
				\SetWidth{1.0}
				\SetColor{Black}
				\Arc(350,-118)(60.803,144,504)
				\Line(225,-190)(297,-149)
				\Line(402,-149)(477,-190)
				\Arc(350,-15)(19.799,135,495)
				\Line(350,26)(350,6)
				\Line(350,-36)(350,-57)
				\Arc[clock](334.625,-15)(15.375,77.32,-77.32)
				\end{picture}}
			&
			\resizebox{7cm}{!}{\begin{picture}(254,218) (224,-191)
				\SetWidth{1.0}
				\SetColor{Black}
				\Arc(261,-169)(19.849,139,499)
				\Line(350,26)(350,-57)
				\Arc(350,-118)(60.803,144,504)
				\Line(225,-190)(244,-179)
				\Line(280,-159)(297,-149)
				\Arc(439,-169)(19.849,139,499)
				\Line(420,-159)(403,-149)
				\Line(477,-190)(457,-179)
				\end{picture}}
			&
			\resizebox{7cm}{!}{\begin{picture}(254,218) (224,-191)
				\SetWidth{1.0}
				\SetColor{Black}
				\Arc(350,-118)(45,143,503)
				\Line(225,-190)(312,-143)
				\Line(387,-143)(477,-190)
				\Arc(350,-42)(15.556,135,495)
				\Arc(350,-1)(15.556,135,495)
				\Line(350,26)(350,15)
				\Line(350,-17)(350,-26)
				\Line(350,-58)(350,-72)
				\end{picture}}
			&
			\resizebox{7cm}{!}{\begin{picture}(254,218) (224,-191)
				\SetWidth{1.0}
				\SetColor{Black}
				\Arc(287,-154)(34.886,153,513)
				\Line(350,26)(350,-11)
				\Line(350,-118)(477,-190)
				\Arc(350,-46)(34.67,147,507)
				\Line(316,-46)(384,-46)
				\Line(350,-81)(350,-118)
				\Line(225,-190)(257,-171)
				\Line(317,-137)(350,-118)
				\end{picture}}
			\\
			{\Huge $g^{ijk}_{(3\epsilon)}$}
			&
			{\Huge $g^{ijk}_{(3\zeta)}$}
			&
			{\Huge $g^{ijk}_{(3\eta)}$}
			&	
			{\Huge $g^{ijk}_{(3\kappa)}$}
			\\
			&
			&
			&
			\\
			\resizebox{7cm}{!}{\begin{picture}(254,218) (224,-191)
				\SetWidth{1.0}
				\SetColor{Black}
				\Arc(287,-154)(34.886,153,513)
				\Line(350,26)(350,-11)
				\Line(350,-118)(477,-190)
				\Arc(350,-46)(34.67,147,507)
				\Line(350,-81)(350,-118)
				\Line(225,-190)(257,-171)
				\Line(317,-137)(350,-118)
				\Arc[clock](323.636,-46)(26.364,80.473,-80.473)
				\end{picture}}
			&
			\resizebox{7cm}{!}{\begin{picture}(255,218) (223,-192)
				\SetWidth{1.0}
				\SetColor{Black}
				\Arc(287,-154)(19.416,125,485)
				\Line(350,-118)(477,-190)
				\Arc(350,-12)(20,127,487)
				\Arc(350,-72)(19.416,125,485)
				\Line(350,25)(350,8)
				\Line(350,-32)(350,-52)
				\Line(350,-92)(350,-118)
				\Line(224,-191)(270,-164)
				\Line(304,-145)(350,-118)
				\end{picture}}
			&
			\resizebox{7cm}{!}{\begin{picture}(254,217) (224,-192)
				\SetWidth{1.0}
				\SetColor{Black}
				\Line(350,-119)(477,-191)
				\Arc(350,-2)(15,127,487)
				\Arc(350,-93)(15,127,487)
				\Line(350,24)(350,14)
				\Line(350,-109)(350,-119)
				\Line(350,-119)(225,-191)
				\Arc(350,-47)(15,127,487)
				\Line(350,-18)(350,-32)
				\Line(350,-63)(350,-78)
				\end{picture}}
			&
			\\
			{\Huge $g^{ijk}_{(3\lambda)}$}
			&
			{\Huge $g^{ijk}_{(3\mu)}$}
			&
			{\Huge $g^{ijk}_{(3\nu)}$}
			&
		\end{tabular*}
	}
	\caption{Three-loop 1PR structures arising from coupling redefinitions}
	\label{onepr}	
\end{table}

and

\begin{align}
\delta c_{(3D)}&=-\tfrac13\epsilon_1-2\epsilon_6+4c_{(2b)}\delta_2-2c_{(2B)}\delta_1
+4\delta_1\delta_2+\tfrac13\delta_1^2,\nn
\delta c_{(3E)}&=-\tfrac16\epsilon_1-2\epsilon_6
+2c_{(2b)}\delta_2-2c_{(2B)}\delta_1+\tfrac14\delta_1^2+4\delta_1\delta_2,\nn
\delta c_{(3F)}&=-\tfrac16 \epsilon_2+\tfrac16\epsilon_6+2c_{(2c)}\delta_2-2c_{(2B)}\delta_2
+4\delta_2^2+\tfrac13\delta_1\delta_2,\nn
\delta c_{(3G)}=\delta c_{(3G')}&=-\tfrac16 \epsilon_2+\tfrac13\epsilon_6-\epsilon_7-\tfrac16\epsilon_5\nn
&~~~+2c_{(2c)}\delta_2-4c_{(2B)}\delta_2-c_{(2C)}\delta_1+\tfrac12\delta_1\delta_2+8\delta_2^2,\nn
\delta c_{(3H)}&=-\tfrac13\epsilon_6-2\epsilon_7+4c_{(2B)}\delta_2-2c_{(2C)}\delta_1,\nn
\delta c_{(3I)}&=-\tfrac16\epsilon_3+2c_{(2d)}\delta_2,\nn
\delta c_{(3J)}&=-\tfrac16\epsilon_4+\tfrac16\epsilon_7-2c_{(2C)}\delta_2
-\tfrac16\delta_2^2 ,\nn
\delta c_{(3L)}&=\tfrac13\epsilon_7-\tfrac13\epsilon_8-4c_{(2C)}\delta_2
-\tfrac16\delta_2^2,
\label{reddefsb}
\end{align}

and also for the 1PR structures

\begin{align}
\delta c_{(3\alpha)}=-2\epsilon_5+2\delta_1\delta_2,\quad
 \delta c_{(3\beta)}&=\tfrac16 \epsilon_5-\tfrac16\delta_1\delta_2,\quad
\delta c_{(3\gamma)}=2\epsilon_4+\tfrac13\epsilon_5-\tfrac13\delta_1\delta_2
-2\delta_2^2,\nn
\delta c_{(3\delta)}=-2\epsilon_5+2\delta_1\delta_2,\quad
\delta c_{(3\epsilon)}&=\tfrac13\epsilon_5-\tfrac13\delta_1\delta_2,\quad
 \delta c_{(3\zeta)}=-\epsilon_4+\delta_2^2,\nn
\delta c_{(3\eta)}=\tfrac16\epsilon_5-\tfrac16\delta_1\delta_2,\quad
\delta c_{(3\kappa)}&=-4\epsilon_4+4\delta_2^2,\quad
\delta c_{(3\lambda)}=\tfrac23\epsilon_4-\tfrac23\delta_2^2,\nn
\delta c_{(3\mu)}=\tfrac13\epsilon_4-\tfrac13\delta_2^2,&\quad
\delta c_{(3\nu)}=\tfrac16\epsilon_8-\tfrac14\delta_2^2,\nn
\delta c_{(3\rho)}&=-\tfrac{1}{12}\epsilon_5+\tfrac{1}{12}\epsilon_6-2\epsilon_8
-c_{(2B)}\delta_2+4\delta_2^2+\tfrac{1}{12}\delta_1\delta_2,\nn
\delta c_{(3\sigma)}&=-\tfrac{1}{12}\epsilon_5-\tfrac{1}{12}\epsilon_6-2\epsilon_8
+c_{(2B)}\delta_2+2\delta_2^2+\tfrac{1}{12}\delta_1\delta_2,\nn
\delta c_{(3\tau)}&=-\tfrac16\epsilon_4+\tfrac{1}{12}\epsilon_7+\tfrac13\epsilon_8
-c_{(2C)}\delta_2-\tfrac12\delta_2^2\nn
\delta c_{(3\chi)}&=-\tfrac16\epsilon_4-\tfrac{1}{12}\epsilon_7+\tfrac13\epsilon_8
+c_{(2C)}\delta_2-\tfrac16\delta_2^2.
\label{reddefs}
\end{align}
One may now verify that the consistency conditions Eqs.~\eqref{conthree} are invariant under
the coupling redefinitions of Eq.~\eqref{redefth}; and so is Eq.~\eqref{conthreea}, provided we assume that $c_{(3G)}^{v}$ is scheme-independent, as is natural for a quantity appearing for the first time at this loop order.
 Finally, it is interesting to note that the condition required for symmetry of $T^{(4)}_{IJ}$, Eq.~\eqref{symcons}, is also invariant under these transformations so that a symmetric $T_{IJ}$ can be obtained in any renormalisation scheme at this order.

The coefficients $c_{(3\alpha)}$--$c_{(3\nu)}$ corresponding to 1PR diagrams, and $c_{(3\rho)}$--$c_{(3\chi)}$ 
corresponding to 1PR contributions to the anomalous dimension, are of course zero in $\MSbar$. 
It would be natural to restrict ourselves to transformations which take us to other schemes with the same property;
which indeed one might expect to be shared by any well-defined diagrammatic renormalisation scheme. 
Now it is clear from Eq.~\eqref{reddefs} that any scheme change such that 
\be
 \epsilon_4=\delta_2^2,\quad \epsilon_5=\delta_1\delta_2,\quad\epsilon_8=\tfrac32\delta_2^2,
\label{epsfour}
\ee
will preserve the vanishing of $c_{(3\alpha)}$--$c_{(3\nu)}$, but will then inevitably give
\begin{align}
 \delta c_{(3\rho)}
=-\delta c_{(3\sigma)}&=\tfrac{1}{12}\epsilon_6+\delta_2^2-c_{(2B)}\delta_2,\nn
\delta c_{(3\tau)}=-\delta c_{(3\chi)}&=\tfrac{1}{12}\epsilon_7-\tfrac16\delta_2^2-c_{(2C)}\delta_2.
\label{redcon}
\end{align}
The values of $c_{(3\rho)}$, $c_{(3\sigma)}$ are opposite in sign, but not manifestly zero; likewise $c_{(3\tau)}$, $c_{(3\chi)}$. Of course one could decide to choose $\epsilon_6$, $\epsilon_7$ so as to ensure the vanishing of 
 $c_{(3\rho)}$--$c_{(3\chi)}$ in Eq.~\eqref{redcon}, but this seems rather restrictive and would not necessarily 
correspond to a natural diagrammatic renormalisation prescription. We shall therefore approach the issue from 
a different angle and consider in some detail the example of MOM, which certainly is
defined by a diagrammatic prescription as described in detail in Sect 3. 
 
The MOM and $\MSbar$ results first part company at two loops. The two-loop MOM $\beta$-function coefficients
may readily be computed directly using the methods described in Sect.~3. They are
given by taking in Eq.~\eqref{betatwo}
\be
c_{(2B)}^{\MOMs}=\tfrac{1}{36}-\tfrac{2}{81}\pi^2+\tfrac{1}{27}\psi'(\tfrac13),
\quad
c_{(2c)}^{\MOMs}=\tfrac{1}{8}+\tfrac{2}{81}\pi^2-\tfrac{1}{27}\psi'(\tfrac13),
\label{cmomtwo}
\ee
where $\psi(z)$ is the Euler $\psi$-function defined by
\be
\psi(z)=\frac{d}{dz}\ln\Gamma(z),
\ee
the other two-loop $\beta$-function coefficients remaining unchanged. Alternatively, we may compute the two-loop MOM
 $\beta$-function coefficients by effecting 
the appropriate scheme change as described above. 
Comparing Eqs.~\eqref{cmstwo}, \eqref{cmomtwo}, we simply require to take in Eq.~\eqref{redone}
\be
\Delta=\tfrac{1}{6} +\tfrac{4}{27}\pi^2-\tfrac{2}{9}\psi'(\tfrac13)
\ee
to effect the change to the MOM values.
Of course this does not specify $\delta_1$, $\delta_2$ in Eq.~\eqref{redone} uniquely.
However, calculating the coupling redefinition required for the change from $\MSbar$ to MOM as described in Section 3, we find that the change to MOM corresponds to taking
\be
\delta_1= \tfrac32+\tfrac{4}{27}\pi^2-\tfrac{2}{9}\psi'(\tfrac13), \quad \delta_2=-\tfrac{1}{9},
\label{redonemom}
\ee
and it is easy to check that Eqs.~\eqref{redone}, \eqref{redonemom} are compatible with the difference 
between Eqs.~\eqref{cmstwo} and \eqref{cmomtwo}.

The transformations required to change 
scheme from $\MSbar$ to MOM at the three-loop level may similarly be calculated 
using the methods described in Sect.~3. They are given by taking in Eq.~\eqref{redefth}
\begin{align}
\epsilon_1&= \tfrac{51}{32} + \tfrac{11}{54}\pi^2 - \tfrac{11}{36}\psi'\left(\tfrac13\right),\nn
\epsilon_2&=  - \tfrac{703}{1728} - \tfrac{41}{972}\pi^2 + \tfrac{41}{648}\psi'\left(\tfrac13\right),\nn
\epsilon_3&= \tfrac{59}{48} - \tfrac12\zeta(3) - \tfrac{7}{27}\pi^2 + \tfrac{1}{144}\ln(3)^2\sqrt3\pi\nn
&~~~- \tfrac{1}{12}\ln(3)\sqrt3\pi - \tfrac{29}{3888}\sqrt3\pi^3
+ 3s_2\left(\tfrac{\pi}{6}\right) - 6s_2\left(\tfrac{\pi}{2}\right)\nn
&~~~ - 5s_3\left(\tfrac{\pi}{6}\right)
+ 4s_3\left(\tfrac{\pi}{2}\right)+ \tfrac{7}{18}\psi'\left(\tfrac13\right),\nn
&\epsilon_6 =- \tfrac{215}{864},\quad
\epsilon_7 =  \tfrac{791}{10368},
\label{expred}
\end{align}
together with $\epsilon_4$, $\epsilon_5$ and $\epsilon_8$ as given in Eqs.~\eqref{epsfour}, with 
$\delta_{1,2}$ as defined in Eq.~\eqref{redonemom}.  The explicit MOM  results for the three-loop
$\beta$-function coefficients obtained by combining Eqs.~\eqref{betthree}, {\eqref{betthreea}, \eqref{reddefsa}, \eqref{reddefsb} are somewhat lengthy and are postponed to Appendix A.

Unfortunately the values of $\epsilon_6$, $\epsilon_7$ shown in Eq.~\eqref{expred}, do not correspond to the 
vanishing of $\delta c_{(3\rho)}$, $\delta c_{(3\sigma)}$, $\delta c_{(3\tau)}$, $\delta c_{(3\chi)}$ in 
Eq.~\eqref{redcon}. The scheme transformation therefore predicts non-vanishing MOM $\beta$-function contributions 
from these 1PR anomalous dimension structures, which seems somewhat counter-intuitive.    
Indeed, after a careful direct calculation using the standard definition of MOM given in Sect 3, 
and taking account of the fact that the relation between $\beta$-function coefficients and renormalisation 
constants is less trivial in MOM than in $\MSbar$, we obtain 
$c^{\MOMs}_{(3\chi)}=c^{\MOMs}_{(3\tau)}=0$. It seems likely that the same applies to $c^{\MOMs}_{(3\rho)}$, 
$c^{\MOMs}_{(3\sigma)}$. We therefore have an apparent inconsistency between the MOM values of
$c_{(3\rho)}\ldots c_{(3\chi)}$ obtained by the coupling redefinition process from $\MSbar$, and those obtained by direct 
calculation. We have checked the anomalous dimension coefficients $c^{\MOMs}_{(3J)}$, 
$c^{\MOMs}_{(3K)}$, $c^{\MOMs}_{(3L)}$ obtained by coupling redefinition as given in Eq.~\eqref{momthreea} by a direct three-loop diagrammatic computation;  it therefore appears likely that the discrepancy only affects the particular 
contriubtions $c_{(3\rho)}\ldots c_{(3\chi)}$ corresponding to 1PR anomalous dimension contributions. We also see from Eq.~\eqref{reddefsb} that this check of $c^{\MOMs}_{(3J)}$, $c^{\MOMs}_{(3L)}$ confirms the value of $\epsilon_7$ and therefore fixes $\delta c_{(3\tau)}=-\delta c_{(3\chi)}\ne0$ in Eq.~\eqref{redcon}. This removes the possibility
that there might be a different choice of $\epsilon_1\ldots\epsilon_7$ in Eq.~\eqref{expred} which would correctly reproduce all the directly-computed MOM coefficients including vanishing values for $c_{(3\rho)}\ldots c_{(3\chi)}$.
It seems that one potential resolution of this problem lies in the use of a hybrid MOM scheme as alluded to 
briefly in Sect. 3, in which the wave-function renormalisation constant is adjusted to give MOM values of 
$c_{(3\rho)}\ldots c_{(3\tau)}$ in agreement with the coupling redefinitions (without altering the values of 
any other $\beta$-function coefficients). 
In six dimensions the issue may appear to be simply a technicality, but it seems probable that similar features arise in the four-dimensional case, which is of more practical interest. We propose to return to the subject in a subsequent article where we shall give full details of the MOM calculations reported here and show how a hybrid scheme can resolve the apparent inconsistencies. We shall also show how our results extend to the four dimensional case. Furthermore, it seems conceivable that a similar adjustment of the wave function renormalisation constant may be required in other schemes in order to match  the results obtained by coupling redefinition to those obtained by direct computation, at least for 1PR anomalous dimension contributions.   




We may now finally compute the MOM values of the coefficients in $A^{(4)}$, $A^{(5)}$. In order for this to be possible we know that the MOM coefficients should satisfy the appropriate consistency conditions derived in Sect.~4. We have already remarked that the two-loop MOM coefficients satisfy Eq.~\eqref{constwo}.
Consequently, starting with $A^{(4)}$, the MOM values of $a_2\ldots a_5$ in Eq.~\eqref{Afour} may be derived by again solving Eq.~\eqref{grad}, but with the MOM values as in Eq.~\eqref{cmomtwo}; or more easily using the fact that $A$ transforms as a scalar under coupling redefinitions,
\be
\overline{A}(\gbar)=A(g).
\label{Atran}
\ee
Using Eqs.~\eqref{redonea}, \eqref{redone}, and \eqref{Athree}, we find for a general one-loop redefinition a leading-order change in
$A$ given by
\be
\delta A^{(4)}=\lambda[\Delta(A_2-\tfrac14A_3)-12\delta_2\beta^{(1)ijk}\beta^{(1)ijk}],
\ee
and consequently for MOM using the values in Eq.~\eqref{redonemom} we may take
\be
a_2^{\MOMs}=a_2+\Delta=\tfrac{1}{24}+\tfrac{4}{27}\pi^2-\tfrac29\psi'(\tfrac13),
\quad
a_3^{\MOMs}=a_3-\tfrac14\Delta=\tfrac{5}{48}-\tfrac{1}{27}\pi^2+\tfrac{1}{18}\psi'(\tfrac13),
\ee
with $a_4$ and $a_5$ unchanged.
Clearly other choices are possible with corresponding adjustments of the value of $\alpha_1$ in Eq.~\eqref{Afour}.

 At the next order, we have mentioned already that the three-loop MOM coefficients in Eqs.~\eqref{momthreea}, \eqref{momthreeb}  satisfy all the consistency conditions in 
Eqs.~\eqref{conthree} and \eqref{conthreea}, provided we take the non-zero MOM values of $c_{(3\rho)}\ldots c_{(3\chi)}$ implied by Eq.~\eqref{redcon} (and listed in Appendix A), and assume that
$c_{(3G)}^{v}$ is scheme-independent, as is natural for a quantity appearing for the first time at this loop order. We have suggested above that these non-zero values correspond to a hybrid MOM scheme\footnote{It is however  interesting to observe that if we take  $c_{(3\rho)}\ldots c_{(3\chi)}$ to be zero, as obtained by direct calculation in the standard MOM scheme, the MOM coefficients satisfy Eq.~\eqref{conthreea} if we take
$c_{(3G)}^{v}=0$.}.
The MOM values of $a_{(1)}^{(5)}\ldots a_{(16)}^{(5)}$ in Eq.~\eqref{athree} may then most easily be derived using Eq.~\eqref{Atran}
and the explicit transformations given by taking Eqs.~\eqref{redonemom}, \eqref{expred} in Eqs.~\eqref{redonea}, \eqref{redefth} respectively. Again, one may also solve the equations using the MOM values of the $\beta$-function coefficients as given in Eqs.~\eqref{cmomtwo}, \eqref{momthreea}, \eqref{momthreeb}; this will yield the same results, up to the freedom expressed in Eq.~\eqref{afree}.

\section{Conclusions}
We have shown that, as in four dimensions, the gradient-flow equation Eq.~\eqref{grad} imposes constraints on the $\beta$-function coefficients, and we have shown that these constraints are satisfied by the explicit results as computed for the  $\MSbar$ and MOM schemes up to three-loop order. We have demonstrated that the tensor $T_{IJ}$ which appears in Eq.~\eqref{grad} may be chosen as symmetric up to this order. We have also shown that for a general scalar theory with an $O(N)$ global invariance, the $\beta$-functions on the right-hand side of Eq.~\eqref{grad} must be replaced at three-loop order in the $\MSbar$ scheme by the generalised ``$B$''-functions, as has also been observed in four dimensions.
It would be useful to extend the analysis of Ref.~\cite{OsbSter} along the lines of Ref.~\cite{OsbJacnew} in order to understand the issues of theories with a global invariance further. This would have the benefit of enabling an explicit calculation of the ``$v$'' term in $B$ (Eq.~\eqref{vval}) and would also allow an understanding of its scheme dependence.


 Finally our analysis of scheme dependence has raised issues concerning the relation of $\MSbar$ and MOM; specifically, the MOM values obtained for certain $\beta$-function coefficients corresponding to 1PR contributions to the anomalous dimension are different depending on whether they are obtained by direct calculation within the standard MOM scheme, or by coupling redefinition from $\MSbar$. We shall discuss this issue further in a subsequent article where we shall show that the apparent discrepancy can be avoided by using a hybrid MOM scheme; we shall also explore similar issues in four-dimensional theories.

\bigskip

{\Large{{\bf{Acknowledgements}}} }\hfil

We are very grateful to Tim Jones and Hugh Osborn for useful conversations and to Hugh Osborn for a careful reading of the manuscript.
This work was supported in part by the STFC under contract ST/G00062X/1, and CP was supported by an STFC studentship.

\appendix

\section{Three-loop MOM results}

For the MOM scheme the three-loop  $\beta$-function coefficients were computed by constructing the appropriate coupling redefinitions as described in Sect. 3. Using the same notation as for the $\MSbar$ scheme, they are given by
\begin{align}
c^{\MOMs}_{(3D)} &=  \tfrac{1}{432} - \tfrac{1}{486} \pi^2
+\tfrac{16}{2187} \pi^4 +\tfrac{1}{324} \psi^\prime(\third)
- \tfrac{16}{729} \psi^\prime(\third) \pi^2
+\tfrac{4}{243} \psi^\prime(\third)^2, \nn
c^{\MOMs}_{(3E)} &=  \tfrac{25}{432} - \tfrac{5}{972} \pi^2
+ \tfrac{4}{729} \pi^4 + \tfrac{5}{648} \psi^\prime(\third)
- \tfrac{4}{243} \psi^\prime(\third) \pi^2
+ \tfrac{1}{81} \psi^\prime(\third)^2, \nonumber \\
c^{\MOMs}_{(3F)} &= \tfrac{5}{5184} + \tfrac{1}{648} \pi^2
- \tfrac{1}{432} \psi^\prime(\third),\nn
c^{\MOMs}_{(3G)} ~&=~ c^{\MOMs}_{(3G')}~=~-\tfrac{7}{432} + \tfrac{31}{5832} \pi^2
- \tfrac{31}{3888} \psi^\prime(\third), \nonumber \\
c^{\MOMs}_{(3H)} &= -\tfrac{107}{2592} + \tfrac{11}{1458} \pi^2
- \tfrac{11}{972} \psi^\prime(\third), \nn
c^{\MOMs}_{(3I)} &= -\tfrac{1}{48} + \tfrac{1}{24} \zeta(3)
+\tfrac{7}{162} \pi^2 - \tfrac{1}{864} \ln(3)^2 \sqrt{3} \pi
+ \tfrac{1}{72} \ln(3) \sqrt{3} \pi + \tfrac{29}{23328} \sqrt{3} \pi^3
\nonumber \\
&~~~ -~ \tfrac{1}{2} s_2(\pisix) + s_2(\pitwo) + \tfrac{5}{6} s_3(\pisix)
- \tfrac{2}{3} s_3(\pitwo) - \tfrac{7}{108} \psi^\prime(\third), \nonumber \\
c^{\MOMs}_{(3J)} &=  \tfrac{103}{31104},\quad
c^{\MOMs}_{(3L)} = \tfrac{85}{15552},
\label{momthreea}
\end{align}
for the anomalous dimension contributions, and
\begin{align}
c^{\MOMs}_{(3g)} &=  \tfrac{1}{8} + \tfrac{1}{9} \pi^2
+ \tfrac{16}{729} \pi^4 - \tfrac{1}{6} \psi^\prime(\third)
- \tfrac{16}{243} \psi^\prime(\third) \pi^2
+ \tfrac{4}{81} \psi^\prime(\third)^2, \nonumber \\
c^{\MOMs}_{(3h)} &=  \tfrac{1}{24} - \tfrac{1}{81} \pi^2
- \tfrac{8}{2187} \pi^4 + \tfrac{1}{54} \psi^\prime(\third)
+ \tfrac{8}{729} \psi^\prime(\third) \pi^2
- \tfrac{2}{243} \psi^\prime(\third)^2, \nn
c^{\MOMs}_{(3i)} ~&= \tfrac{1}{12} - \tfrac{2}{81} \pi^2
- \tfrac{16}{2187} \pi^4 + \tfrac{1}{27} \psi^\prime(\third)
+ \tfrac{16}{729} \psi^\prime(\third) \pi^2
- \tfrac{4}{243} \psi^\prime(\third)^2, \nonumber \\
c^{\MOMs}_{(3j)} &=  \tfrac{3}{16} - \tfrac{5}{486} \pi^2
- \tfrac{16}{2187} \pi^4 + \tfrac{5}{324} \psi^\prime(\third)
+ \tfrac{16}{729} \psi^\prime(\third) \pi^2
-\tfrac{4}{243} \psi^\prime(\third)^2, \nn
c^{\MOMs}_{(3k)} ~&=~ \tfrac{1}{12} + \tfrac{1}{81} \pi^2
- \tfrac{1}{54} \psi^\prime(\third), \nonumber \\
c^{\MOMs}_{(3l)} &=  \tfrac{1}{162} \pi^2
- \tfrac{1}{108} \psi^\prime(\third),\nn
c^{\MOMs}_{(3m)} ~&=~ -\tfrac{1}{16} - \tfrac{31}{2916} \pi^2
+ \tfrac{31}{1944} \psi^\prime(\third), \nonumber \\
c^{\MOMs}_{(3n)} &= -\tfrac{7}{288} - \tfrac{1}{324} \pi^2
+ \tfrac{1}{216} \psi^\prime(\third),\nn
c^{\MOMs}_{(3o)} ~&=~ -\tfrac{7}{288} - \tfrac{1}{324} \pi^2
+\tfrac{1}{216} \psi^\prime(\third), \nonumber \\
c^{\MOMs}_{(3p)} &=  \tfrac{19}{72} - \tfrac{1}{6} \zeta(3)
- \tfrac{7}{81} \pi^2 + \tfrac{1}{432} \ln(3)^2 \sqrt{3} \pi
- \tfrac{1}{36} \ln(3) \sqrt{3} \pi - \tfrac{29}{11664} \sqrt{3} \pi^3
+ s_2(\pisix) \nn
& -~ 2 s_2(\pitwo) - \tfrac{5}{3} s_3(\pisix) + \tfrac{4}{3} s_3(\pitwo)
+\tfrac{7}{54} \psi^\prime(\third), \nonumber \\
c^{\MOMs}_{(3q)} &= -\tfrac{13}{24} +\tfrac{1}{2} \zeta(3)
+ \tfrac{1}{3} \pi^2 - \tfrac{1}{144} \ln(3)^2 \sqrt{3} \pi
+ \tfrac{1}{12} \ln(3) \sqrt{3} \pi + \tfrac{29}{3888} \sqrt{3} \pi^3
- 3 s_2(\pisix) \nonumber \\
&~~~ + 6 s_2(\pitwo) + 5 s_3(\pisix)
- 4 s_3(\pitwo) - \tfrac{1}{2} \psi^\prime(\third),\nonumber \\
c^{\MOMs}_{(3r)} &= -\tfrac{23}{12} + 2 \zeta(3)
+ \tfrac{2}{3} \pi^2 - \tfrac{1}{72} \ln(3)^2 \sqrt{3} \pi
+\tfrac{1}{6} \ln(3) \sqrt{3} \pi + \tfrac{29}{1944} \sqrt{3} \pi^3
- 6 s_2(\pisix) \nonumber \\
&~~~+~ 12 s_2(\pitwo) +10 s_3(\pisix)
- 8s_3(\pitwo) - \psi^\prime(\third), \nonumber \\
c^{\MOMs}_{(3s)} &= -\tfrac{1}{8} - \tfrac{1}{3} \pi^2
+\tfrac{1}{144} \ln(3)^2 \sqrt{3} \pi - \tfrac{1}{12} \ln(3) \sqrt{3} \pi
- \tfrac{29}{3888} \sqrt{3} \pi^3 + 3 s_2(\pisix) - 6 s_2(\pitwo) \nonumber \\
& ~~~-~ 5 s_3(\pisix) + 4 s_3(\pitwo) + \tfrac{1}{2} \psi^\prime(\third),
\label{momthreeb}\end{align}
for the 1PI contributions.
In Eqs.~\eqref{momthreea}, \eqref{momthreeb}, we define
\be
s_n(z)=\frac{1}{\sqrt3}{\cal I}\left[\hbox{Li}_n\left(\frac{e^{iz}}{\sqrt3}\right)\right],
\ee

\begin{table}[h]
	\setlength{\extrarowheight}{0.5cm}
	\setlength{\tabcolsep}{24pt}
	\hspace*{-10.75cm}
	\centering
	\resizebox{6cm}{!}{
		\begin{tabular*}{20cm}{ccccc}
			\begin{picture}(266,206) (197,-152)
			\SetWidth{1.0}
			\SetColor{Black}
			\Line(330,19)(250,-117)
			\Line(330,19)(410,-117)
			\Line(250,-117)(410,-117)
			\Line(330,53)(330,19)
			\Line(250,-117)(230,-151)
			\Line(410,-117)(430,-151)
			\Line[arrow,arrowpos=0.5,arrowlength=5,arrowwidth=2,arrowinset=0.2](343,52)(343,22)
			\Line[arrow,arrowpos=0.5,arrowlength=5,arrowwidth=2,arrowinset=0.2](222,-143)(239,-116)
			\Line[arrow,arrowpos=0.5,arrowlength=5,arrowwidth=2,arrowinset=0.2](437,-141)(421,-117)
			\Text(212,-123)[]{\Huge{\Black{$p$}}}
			\Text(364,39)[]{\Huge{\Black{$r$}}}
			\Text(446,-124)[]{\Huge{\Black{$q$}}}
			\end{picture}
			&
			\begin{picture}(266,215) (197,-152)
			\SetWidth{1.0}
			\SetColor{Black}
			\Line(330,28)(250,-108)
			\Line(330,28)(410,-108)
			\Line(250,-108)(410,-108)
			\Line(330,62)(330,28)
			\Line(250,-108)(230,-142)
			\Line(410,-108)(430,-142)
			\Line[arrow,arrowpos=0.5,arrowlength=5,arrowwidth=2,arrowinset=0.2](343,61)(343,31)
			\Line[arrow,arrowpos=0.5,arrowlength=5,arrowwidth=2,arrowinset=0.2](222,-134)(239,-107)
			\Line[arrow,arrowpos=0.5,arrowlength=5,arrowwidth=2,arrowinset=0.2](437,-132)(421,-108)
			\Text(212,-114)[]{\Huge{\Black{$p$}}}
			\Text(364,48)[]{\Huge{\Black{$r$}}}
			\Text(446,-115)[]{\Huge{\Black{$q$}}}
			\Arc(330,-24)(116,-133.603,-46.397)
			\end{picture}
			&
			\begin{picture}(266,215) (197,-152)
			\SetWidth{1.0}
			\SetColor{Black}
			\Line(330,28)(250,-108)
			\Line(330,28)(410,-108)
			\Line(250,-108)(410,-108)
			\Line(330,62)(330,28)
			\Line(250,-108)(230,-142)
			\Line(410,-108)(430,-142)
			\Line[arrow,arrowpos=0.5,arrowlength=5,arrowwidth=2,arrowinset=0.2](343,61)(343,31)
			\Line[arrow,arrowpos=0.5,arrowlength=5,arrowwidth=2,arrowinset=0.2](222,-134)(239,-107)
			\Line[arrow,arrowpos=0.5,arrowlength=5,arrowwidth=2,arrowinset=0.2](437,-132)(421,-108)
			\Text(212,-114)[]{\Huge{\Black{$p$}}}
			\Text(364,48)[]{\Huge{\Black{$r$}}}
			\Text(446,-115)[]{\Huge{\Black{$q$}}}
			\Arc(330,-24)(116,-133.603,-46.397)
			\Vertex(330,-140){10}
			\end{picture}
			&
			\begin{picture}(266,206) (197,-152)
			\SetWidth{1.0}
			\SetColor{Black}
			\Line(330,19)(250,-117)
			\Line(330,19)(410,-117)
			\Line(250,-117)(410,-117)
			\Line(330,53)(330,-117)
			\Line(250,-117)(230,-151)
			\Line(410,-117)(430,-151)
			\Line[arrow,arrowpos=0.5,arrowlength=5,arrowwidth=2,arrowinset=0.2](343,52)(343,22)
			\Line[arrow,arrowpos=0.5,arrowlength=5,arrowwidth=2,arrowinset=0.2](222,-143)(239,-116)
			\Line[arrow,arrowpos=0.5,arrowlength=5,arrowwidth=2,arrowinset=0.2](437,-141)(421,-117)
			\Text(212,-123)[]{\Huge{\Black{$p$}}}
			\Text(364,39)[]{\Huge{\Black{$r$}}}
			\Text(446,-124)[]{\Huge{\Black{$q$}}}
			\end{picture}
			&
			\begin{picture}(266,206) (197,-152)
			\SetWidth{1.0}
			\SetColor{Black}
			\Line(330,19)(250,-117)
			\Line(330,19)(410,-117)
			\Line(330,53)(330,19)
			\Line(250,-117)(230,-151)
			\Line(410,-117)(430,-151)
			\Line[arrow,arrowpos=0.5,arrowlength=5,arrowwidth=2,arrowinset=0.2](343,52)(343,22)
			\Line[arrow,arrowpos=0.5,arrowlength=5,arrowwidth=2,arrowinset=0.2](222,-143)(239,-116)
			\Line[arrow,arrowpos=0.5,arrowlength=5,arrowwidth=2,arrowinset=0.2](437,-141)(421,-117)
			\Text(212,-123)[]{\Huge{\Black{$p$}}}
			\Text(364,39)[]{\Huge{\Black{$r$}}}
			\Text(446,-124)[]{\Huge{\Black{$q$}}}
			\Line(403,-106)(297,-37)
			\Line(257,-106)(320,-64)
			\Line(338,-53)(363,-37)
			\end{picture}
			\\
			&
			&
			&
			&
			\\
			{\Huge $\mathcal{M}^{(1)}_{31}$}
			&
			{\Huge $\mathcal{M}^{(2)}_{42}$}
			&
			{\Huge $\mathcal{M}^{(2)}_{43}$}
			&	
			{\Huge $\mathcal{M}^{(2)}_{52}$}
			&	
			{\Huge $\mathcal{M}^{(2)}_{61}$}
		\end{tabular*}
	}
	\caption{One- and two-loop master integrals}
	\label{master}	
\end{table}

where $\hbox{Li}_n(z)$ is the polylogarithm function.
The values of the coefficients $c_{(3K)}$, $c_{(3e)}$, $c_{(3f)}$, $c_{(3t)}$, and $c_{(3u)}$ are identical in the two schemes.  As mentioned in the main text, we have independently computed the values of $c^{\MOMs}_{(3J)}$, $c^{\MOMs}_{(3K)}$, $c^{\MOMs}_{(3L)}$ {\it ab initio} and verified that we obtain the same values. Also, the coefficients corresponding to the 1PR anomalous dimension contributions obtained by coupling redefinition are
\begin{align}
c_{(3\rho)}&=-c_{(3\sigma)}=-\tfrac{23}{10368},\nn
c_{(3\tau)}&=-c_{(3\chi)}=\tfrac{61}{41472}.
\end{align}

\section{Master integrals}

In this appendix we record the explicit values of the various one and two loop six-dimensional master integrals which were needed to perform the renormalisation in the MOM scheme. Their derivation is based on the values of the corresponding master integrals in four dimensions, which were given in Ref.~\cite{jag18}, and are reproduced above in Table \ref{master}, where ${\cal M}^{(2)}_{43}$ involves the square of a propagator, denoted by a ``dot''. Using the same notation as Ref.~\cite{jag18} to denote the various graphs, the one-loop triangle master integral is

\begin{eqnarray}
{\cal M}^{(1)}_{31} &&= \frac{1}{2\epsilon}
+ \frac{3}{2} + \frac{4}{27} \pi^2 - \frac{2}{9} \psi^\prime(\third)
\nonumber \\
&&
+~ \left[ \frac{7}{2} + \frac{23}{216} \pi^2 + 4 s_3(\pisix)
- \frac{2}{9} \psi^\prime(\third)
- \frac{35\pi^3}{324 \sqrt{3}} - \frac{\ln(3)^2 \pi}{12 \sqrt{3}} \right]
\epsilon ~+~ O(\epsilon^2) ~.~
\end{eqnarray}

At two loops we have
\begin{eqnarray}
{\cal M}^{(2)}_{42} &&=
\left[ \frac{1}{144 \epsilon^2} + \frac{65}{1728 \epsilon}
+ \frac{1}{62208} [ 96 \psi^\prime(\third) - 136 \pi^2 + 8499 ]
\right. \nonumber \\
&& \left. ~
+~ [ 288 \sqrt{3} \ln(3)^2 \pi - 1728 \sqrt{3} \ln(3) \pi
+ 32 \sqrt{3} \pi^3 + 12000 \psi^\prime(\third)
\right. \nonumber \\
&& \left. ~~~~~~
+~ 62208 s_2(\pisix)
- 124416 s_2(\pitwo) - 124416 s_3(\pisix) + 82944 s_3(\pitwo)
\right. \nonumber \\
&& \left. ~~~~~~
-~ 12680 \pi^2 - 34560 \zeta(3) + 318363 ]
\frac{\epsilon}{746496} ~+~ O(\epsilon^2) \right] \mu^4 \nonumber \\
{\cal M}^{(2)}_{43} &&=
\left[ \frac{1}{24 \epsilon^2} + \frac{5}{16 \epsilon}
+ \frac{1}{2592} [ - 120 \psi^\prime(\third) + 62 \pi^2 + 3915 ]
\right. \nonumber \\
&& \left. ~
+ [ - 72 \sqrt{3} \ln(3)^2 \pi - 216 \sqrt{3} \ln(3) \pi
- 136 \sqrt{3} \pi^3 - 2664 \psi^\prime(\third)
\right. \nonumber \\
&& \left. ~~~~~
+~ 7776 s_2(\pisix)
- 15552 s_2(\pitwo) + 10368 s_3(\pitwo) + 966 \pi^2 - 4320 \zeta(3)
\right. \nonumber \\
&& \left. ~~~~~
+~ 93555 ] \frac{\epsilon}{15552} ~+~ O(\epsilon^2) \right] \mu^2 \nonumber \\
{\cal M}^{(2)}_{52} &&=
\left[ \frac{1}{12 \epsilon^2} + \frac{25}{48 \epsilon}
+ \frac{205}{96} + \frac{7}{648} \pi^2
- \frac{1}{27} \psi^\prime(\third) ~+~ O(\epsilon) \right] \mu^2 \nonumber \\
{\cal M}^{(2)}_{61} &&=
\frac{1}{4 \epsilon} \nonumber \\
&&
+~ \left[ \frac{59}{24} - \zeta(3) - \frac{28}{54} \pi^2 + 6 s_2(\pisix)
- 12 s_2(\pitwo)
- 10 s_3(\pisix) + 8 s_3(\pitwo)
\right. \nonumber \\
&& \left. ~~~~~
+~ \frac{7}{9} \psi^\prime(\third)
- \frac{29}{1944} \sqrt{3} \pi^3 - \frac{1}{6} \sqrt{3} \ln(3) \pi
+ \frac{1}{72} \sqrt{3} \ln(3)^2 \pi \right] ~+~ O(\epsilon) ~.~~
\end{eqnarray}
The values for the remaining two-loop masters corresponding to ${\cal M}^{(1)}_{21}$,
${\cal M}^{(2)}_{31}$, ${\cal M}^{(2)}_{41}$ and ${\cal M}^{(2)}_{51}$ of Ref.~\cite{jag18} are
trivial to construct as they correspond to products of one-loop masters, or the
two-loop sunset graph in the case of ${\cal M}^{(2)}_{31}$. We note that the harmonic polylogarithms are based on the theory of cyclotomic polynomials\cite{blum}

\end{document}